\documentclass[useAMS,usenatbib,usegraphicx]{mn2e}

\usepackage{lscape} %to use \begin{landscape}
\usepackage{placeins}

\title[Assembly of stellar haloes in massive ETGs]{The cosmic assembly of stellar haloes in massive Early-Type Galaxies}%{Unprecedented  view of the assembly of early-type massive galaxies at z$\sim$0.65 in the HUDF12}
\author[F. Buitrago et al.]{Fernando Buitrago$^{1,2,3}$ \thanks{E-mail: fbuitrago@oal.ul.pt}, Ignacio Trujillo$^{4,5}$, Emma Curtis-Lake$^{1,6}$, Mireia Montes$^{7}$, 
\newauthor Andrew P. Cooper$^{8}$, Victoria A. Bruce$^{1}$, , Pablo G. P\'erez-Gonz\'alez$^{9}$, Michele Cirasuolo$^{10}$
\\
\\
$^{1}$SUPA\thanks{Scottish Universities Physics Alliance}, Institute for Astronomy, University of Edinburgh, Royal Observatory, Edinburgh, EH9 3HJ, U.K. \\
$^{2}$Instituto de Astrof\'{\i}sica e Ci\^{e}ncias do Espa\c{c}o, Universidade de Lisboa, OAL, Tapada da Ajuda, PT1349-018 Lisbon, Portugal\\
$^{3}$Departamento de F\'{i}sica, Faculdade de Ci\^{e}ncias, Universidade de Lisboa, Edif\'{i}cio C8, Campo Grande, PT1749-016 Lisbon, Portugal\\
$^{4}$Instituto de Astrof\'{i}sica de Canarias, V\'{i}a L\'{a}ctea $s\backslash n$, 38200 La Laguna, Tenerife, Spain \\
$^{5}$Departamento de Astrof\'{i}sica, Universidad de La Laguna, E-38205, La Laguna, Tenerife, Spain  \\
$^{6}$Sorbonne Universit\'es, UPMC-CNRS, UMR7095, Institut d'Astrophysique de Paris, F-75014, Paris, France \\
$^{7}$Department of Astronomy, Yale University, New Haven, CT 06511, USA \\
$^{8}$Institute for Computational Cosmology, Durham, UK \\
$^{9}$Departamento de Astrof\'isica, Facultad de CC. F\'isicas, Universidad Complutense de Madrid, E-28040 Madrid, Spain \\
$^{10}$European Southern Observatory Karl-Schwarzschild-Strasse 2, D-85748 Garching bei Muenchen, Germany
 }

\newcommand {\gtrsim} {\ {\raise-.5ex\hbox{$\buildrel>\over\sim$}}\ }
\newcommand {\lesssim} {\ {\raise-.5ex\hbox{$\buildrel<\over\sim$}}\ } 
\begin{document}

% \date{Accepted ; Received ; in original form}
% \pagerange{\pageref{firstpage}--\pageref{lastpage}} \pubyear{2011}

\maketitle

\label{firstpage}
% \clearpage

\begin{abstract}
Using the exquisite depth of the Hubble Ultra Deep Field (HUDF12 programme) dataset, 
we explore the ongoing assembly of the outermost regions of the most massive galaxies 
($\rm M_{\rm stellar}\geq$ 5$\times$10$^{10}$ M$_{\odot}$) at $z \leq$ 1. 
The outskirts of massive objects, particularly Early-Types Galaxies (ETGs),  are expected to suffer a dramatic 
transformation across cosmic time due to continuous accretion of small galaxies.
HUDF imaging allows us to study this process at intermediate redshifts in 6 massive galaxies,
exploring the individual surface brightness profiles out to $\sim$25 effective radii.
We find that 5-20\% of the total stellar mass for the galaxies in our sample is contained within 10 $< R <$ 50 kpc.
These values are in close agreement with numerical simulations, and higher than those reported for local late-type galaxies ($\lesssim$5\%).
The fraction of stellar mass stored in the outer envelopes/haloes of Massive Early-Type Galaxies 
increases with decreasing redshift, being 28.7\% at $< z > =$ 0.1, 15.1\% at $< z > =$ 0.65 and 3.5\% at $< z > =$ 2.
The fraction of mass in diffuse features linked with ongoing 
minor merger events is $>$ 1-2\%, very similar to predictions based on observed close
pair counts.
Therefore, the results for our small albeit meaningful sample suggest that the size and mass growth of the most massive galaxies have been solely driven by 
minor and major merging from $z =$ 1 to today.
\end{abstract}


\section{Introduction}
\label{sec:intro}

There is ample evidence that the most massive galaxies of the Universe have grown dramatically
in size since $z =$ 3 \citep[][to name but a few]{Daddi05, Trujillo06a, Trujillo06b, Toft07, Cimatti08, Buitrago08, Damjanov09, vanDokkum10,
Cassata11, Bell12, Bruce12, Huertas-Company13}. Early-Type Galaxies (ETGs) --selected by their
morphological classification, or through a proxy like colours or quiescent star formation-- are those that display the most extreme evolution \citep[with sizes $\sim$5 times smaller on average, at a given stellar mass, than their local Universe counterparts;][]{Trujillo07, Buitrago08,vanderWel14}. 

Theoretically, massive galaxies are predicted to undergo a two-phase formation process whereby
there is a initial very rapid and dissipative gas collapse at high-z where most of the in-situ stars originate
(\citet{Khochfar06,Oser10,Ceverino15,Zolotov15,Wellons16}, see observations in \citet{Ricciardelli10,Barro13,Huang13,Williams14} as well).
The next stage must be a combination of major and minor mergers \citep{Bezanson09,Hopkins09b,Ferreras14,Xie15}, 
as these processes best reproduce the observed tight scatter in the size-mass relation of massive galaxies, 
and can account for the only mild mass increase in these systems from high redshift to the present day. 
In this context, some growth is also expected from residual star formation \citep{Perez-Gonzalez08b, Fumagalli14}.
As a consequence, galaxies progressively build up their outer parts (aka galactic outskirts or outer stellar envelopes) 
and thus grow in an inside-out fashion \citep{vanDokkum10,Trujillo11,Buitrago13}.

%Many studies conducted for massive galaxies ($\rm M_{\rm stellar}\geq 5\times10^{10} - 10^{11} \rm M_{\odot}$)
%point out a continuous build-up of their external parts (aka galactic outskirts)
%as the reason for the dramatic changes these objects undergo with redshift \citep{VanDokkum10,Carrasco10}.
%As a result, effective radii and S\'ersic indices will show a progressive increase
%as cosmic time advances \citep{Buitrago13}.

Many observational problems prevent us from directly testing the aforementioned scenario.
First, the outskirts of galaxies are intrinsically the faintest parts of these systems.
Secondly, surface brightness dimming rises very steeply by $(1+z)^3$ \citep[see][]{Giavalisco1996, Ribeiro16}.
Therefore, if these studies are extremely challenging in the local Universe, conducting
them at high redshift has been regarded as unfeasible. 

%In order to probe this evolution, one has to deal with very faint surface brightness
%outer galaxy parts, and this is specially challenging at high-z where cosmological dimming
%becomes rapidly -- rising by a rampant $(1+z)^{-4}$ -- an issue. This is even more problematic for
%spheroid-like/early-type/``red \& dead" objects \citep{Trujillo07,Buitrago08,Cassata11,Bell12,Huertas-Company13},
%because their luminosity profiles are dominated by extended outer wings in comparison with
%more exponential disk-like galaxies, and those features are prone to become buried
%under the noise level of common extragalactic images.

Various techniques have been applied in order to overcome these hurdles in the local Universe 
and to extract the information enclosed in the outer regions of massive galaxies. These include: stacking \citep{Zibetti04a,Tal11,LaBarbera12,D'Souza14},
deep photometric studies \citep{Zibetti04b,Atkinson13,vanDokkum14,Duc15,Trujillo16}, very deep spectroscopic analyses \citep{Coccato10} or stellar counts \citep{Crnojevic13,Rejkuba14}. In doing so we have learned
that $\sim$70\% of the nearby massive ETGs 
show features indicative of mergers or the tidal disruption of less massive companions \citep{vanDokkum05,Tal09,Kaviraj10}. The observed features, such as shells or
tidal tails, are red, smooth and extended (sometimes $>$ 50 kpc). 
This has led to an overall consensus that these galaxies are assembled via mergers involving gas-poor and bulge-dominated systems.

%Ideally, we would like to obtain similar measurements at higher redshifts, and
%especially for massive galaxies where their size and morphological changes provide us with compelling
%evidence about an accelerated evolution in comparison with lower mass systems \citep{Buitrago13}. 
The new observations of the Hubble Ultra Deep Field (HUDF), in particular the HUDF12 programme \citep{Ellis13,Koekemoer13},
have opened up the possibility of exploring galaxies to an unprecedented level of detail
(5$\sigma$ limiting magnitude $\sim$30 AB mag). 
The extraordinary depth and resolution of these observations,
combined with the fact that HUDF12 is the only HUDF programme which preserves the galaxy
extended envelopes/haloes, enable us to study galaxy surface brightness profiles down to 31 mag arcsec$^{-2}$ 
 or 25 effective radii ($r_{e}$) for the galaxies in our sample, sometimes
reaching $\sim$100 kpc in galactocentric distance.

In the present paper we perform
an investigation on the nature of the galaxy outskirts at large galactocentric distances
in these ETGs, trying to understand their observables (e.g. percentage
of light and mass with respect to the central parts, colours, mass profiles),
focusing our study on constraining the mass assembly of massive galaxies,
giving the first, direct measurement of the mass growth by ongoing mergers.
%A companion paper (Buitrago et al. in prep.) will show
%the impact of our results for the mass-size relation.

The structure of the paper is as follows: Sections \ref{sec:data}, \ref{sec:sample} and \ref{sec:analysis}
present the data, the sample and the analysis respectively. Section \ref{sec:results} 
shows the several tests we carried out for describing the stellar haloes in
our sample of massive galaxies and finally, Section \ref{sec:conclusions} delivers
our summary and conclusions. Hereafter, we adopt
 a cosmology with $\Omega_m$=0.3, $\Omega_\Lambda$=0.7 and H$_0$=70 kms$^{-1}$
 Mpc$^{-1}$. We use a \citet{Chabrier03} Initial Mass Function (IMF), unless otherwise stated. 
Magnitudes are provided in the AB system \citep{Oke1983}.

\begin{figure*}
% \begin{center} 
\vspace{0.5cm}
\hspace{2cm}
\rotatebox{0}{
\includegraphics[angle=0,width=1.0\linewidth]{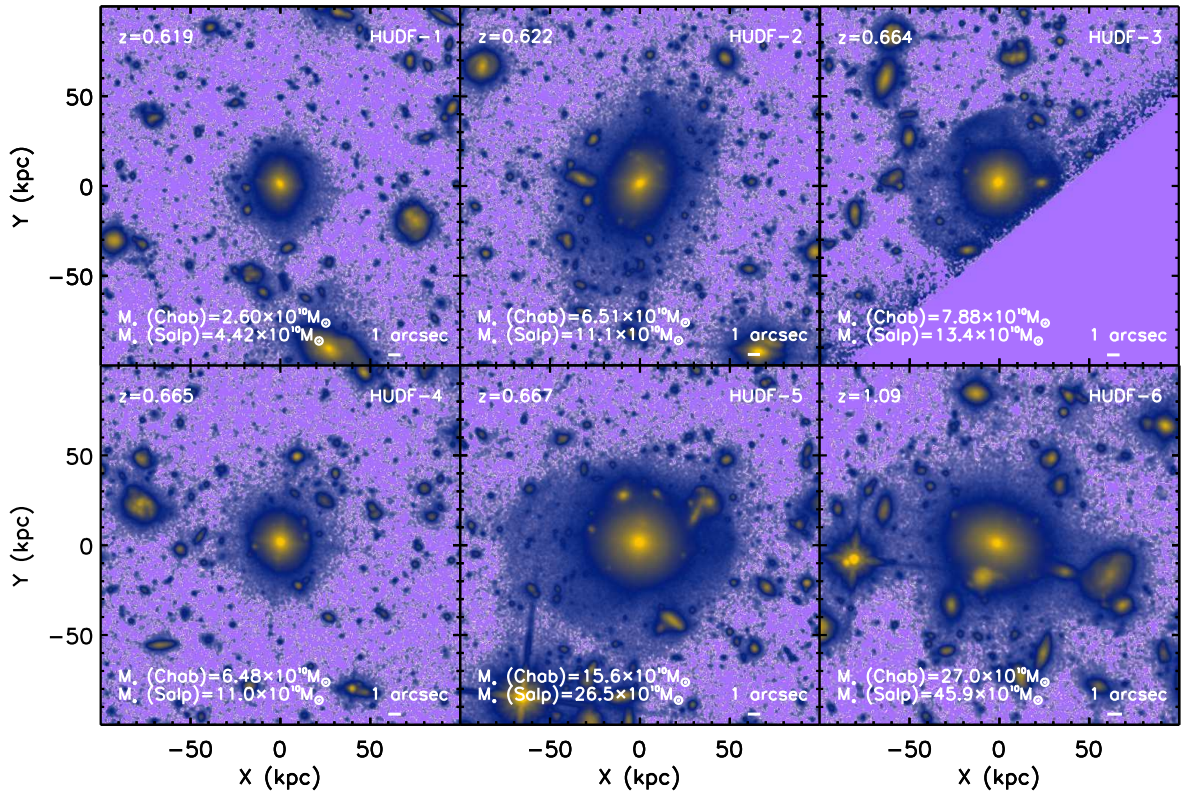}}
% \end{center}
\vspace{-0.5cm}

\caption{Montage with the HUDF12 WFC3 images for our sample of massive ETGs, also showing their spectroscopic
redshifts and photometric masses. These are the stacked HST NIR images, and the colour
palette ranges from 18 to 30 mag arcsec$^{2}$. The superb WFC3 resolution
(approximately 0.18 arcsec, $\sim$1.25 kpc at $<z>$ = 0.65, the median redshift of our observations)
allow us to see the huge stellar envelopes for these objects, apart from broad
fans of stars or shells (for HUDF-3 and HUDF-5) and other asymmetries.
It is also striking the presence of so many potential satellites, which may 
contribute to the size increase of the massive objects via minor merging.}

\label{fig:mosaicluminanceIR}

\end{figure*}

\begin{figure*}
% \begin{center} 
\vspace{0.5cm}
\hspace{2cm}
\rotatebox{0}{
\includegraphics[angle=0,width=1.0\linewidth]{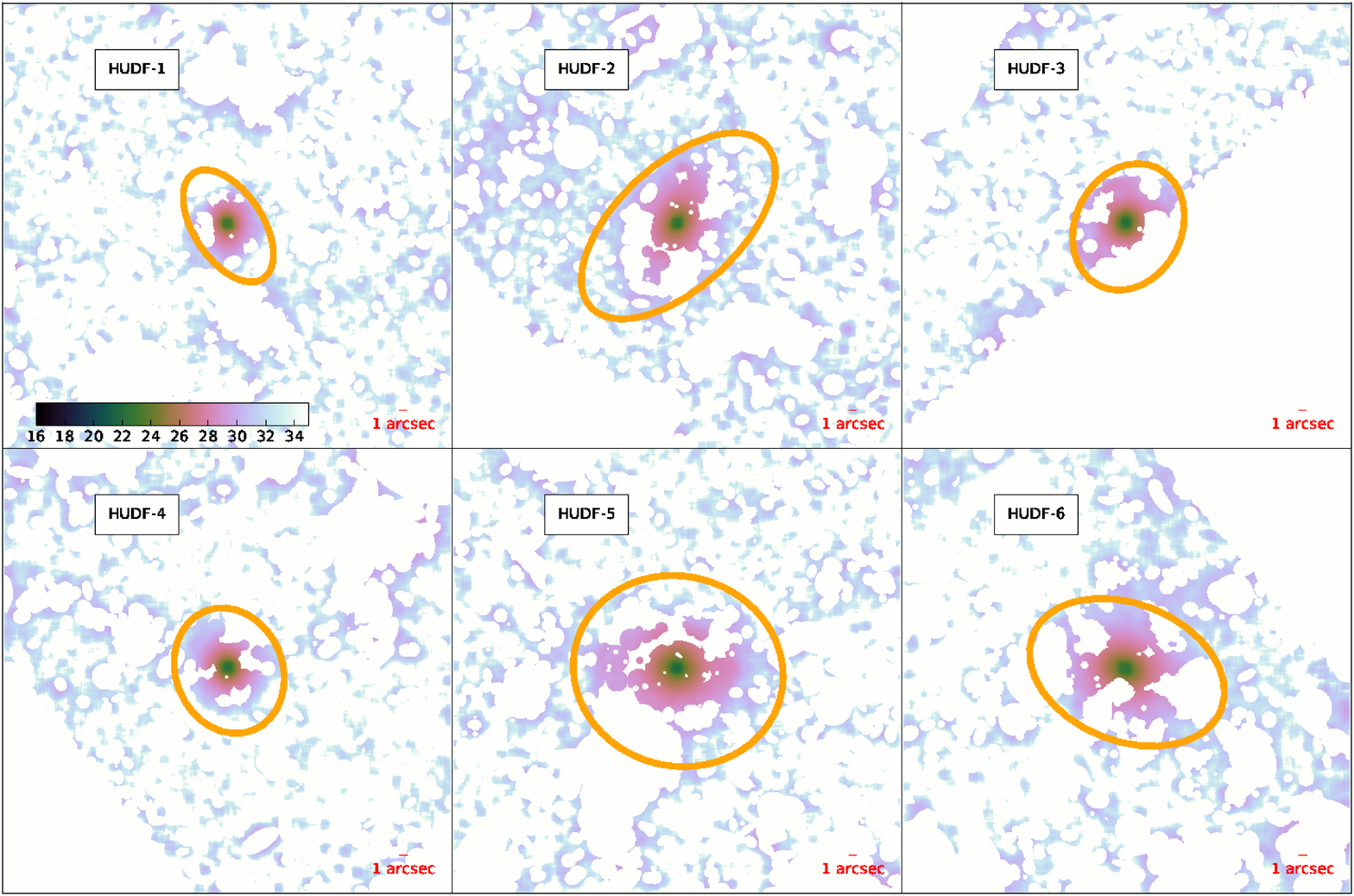}}
% \end{center}
\vspace{-0.5cm}

\caption{Coaddition of all near-infrared WFC3 images. The resulting image has been smoothed by convolving a 2-arcsec standard deviation Gaussian kernel and then our NIR mask of the galaxy neighbours has been overploted. The whole process is done in order to highlight the lowest signal-to-noise features in the image. The colorbar displays the surface brightness in units of mag arcsec$^{-2}$. The golden ellipse shows the extent of our surface brightness analysis in the H-band (reddest band in the NIR).}
\label{fig:masks_nir}

\end{figure*}

\begin{figure*}
% \begin{center} 
\vspace{0.5cm}
\hspace{2cm}
\rotatebox{0}{
\includegraphics[angle=0,width=1.0\linewidth]{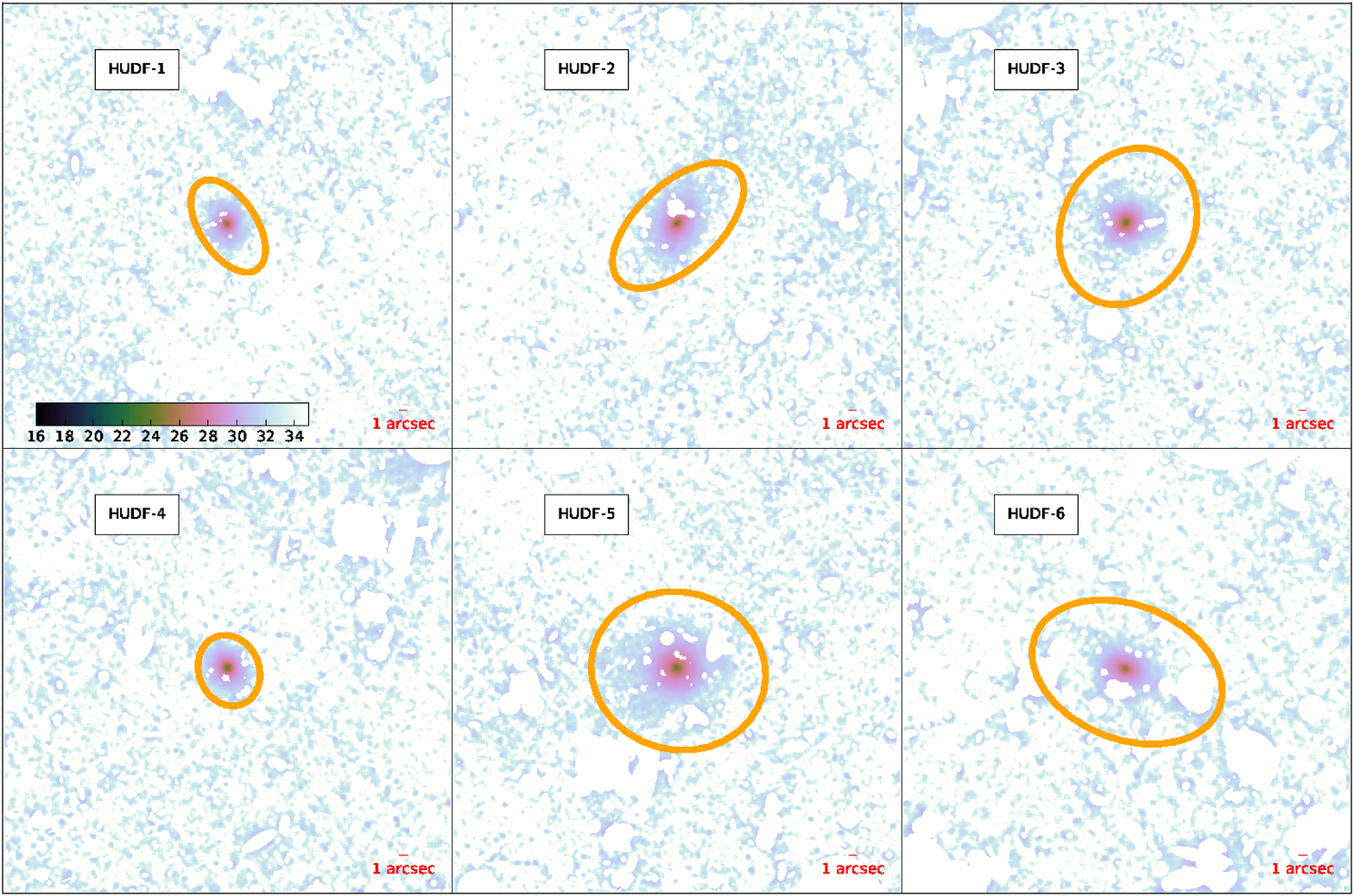}}
% \end{center}
\vspace{-0.5cm}

\caption{Coaddition of all optical ACS images. The resulting image has been smoothed by convolving a 2-arcsec standard deviation Gaussian kernel and then our optical mask of the galaxy neighbours has been overploted. The whole process is done in order to highlight the lowest signal-to-noise features in the image. The colorbar displays the surface brightness in units of mag arcsec$^{-2}$. The golden ellipse shows the extent of our surface brightness analysis in the Z-band (reddest band in the optical).}
\label{fig:masks_opt}

\end{figure*}

\section{The data}
\label{sec:data}

We analyzed the deepest ever HST observations, the Hubble Ultra Deep Field
(HUDF; R.A. = 03:32:39.0, DEC = -~27:47:29.1, J2000). In order to detect extended stellar haloes
around intermediate-redshift galaxies, the best Near-InfraRed (NIR) data available was provided by
the HUDF12\footnote{http://archive.stsci.edu/pub/hlsp/hudf12/} programme \citep{Ellis13,Koekemoer13}. 
This survey combines the images from the HUDF09 programme
\citep[][and references therein]{Bouwens12} with a new 128-orbit campaign
(HST Program ID 12498, PI.: R. Ellis and R. McLure). This translates into
an outstanding improvement of the previous dataset, by enhancing the exposure times
(sometimes even quadrupling them, as for the F105W filter) and adding new imaging in the F140W filter.
Additionally, and key for our purposes, HUDF12 is unique as its HUDF data reduction preserves
the faint wings of extended sources. Finally, in order to obtain the largest multiwavelength
HST coverage, we also make use of the optical ACS observations\footnote{http://archive.stsci.edu/pub/hlsp/udf/acs-wfc/} over the same area \citep{Beckwith06}.
Therefore, we have investigated the area (4.7 arcmin$^{2}$) where WFC3 and ACS observations overlap.
We list the photometric bands, total exposure times and zeropoints in Table \ref{tab:filters}.

We also need to understand whether the level of background fluctuations in these data enables us to characterize very faint surface brightness features.
We conducted a thorough characterization of each science image by placing 25000 random square 1$\times$1 arcsec apertures in empty sky patches,
inferring a surface brightness limit of $\geq$31 mag arcsec$^{-2}$ (only slighty brighter for Z-band) at a 3$\sigma$ level over the background fluctuations in 10$\times$10 arcsec boxes.
The WFC3/IR images have passed multiple checks regarding their sky background properties, especially about persistence
and large-scale flatfield variations. After the applied residual corrections, the sky is flat to within $\sim$1-2$\%$ of mean
sky level, translating into uniform limiting depths ($\leq 0.03$ mag) throughout the images. On the other hand, as it has been already indicated, the ACS programme
targeting the HUDF is prior to the HUDF12 campaign, which is actually an asset to minimize the charge transfer efficiency degradation caused by radiation damage. 
It is to note that careful flatfielding and treatment of the scattered light was undertaken, as described at length in the Section 3.1 in \citet{Beckwith06}.
The final residual flux is also $<$2$\%$ of the sky level.

\section{The sample}
\label{sec:sample}

The criteria for our galaxy selection are the following: ETG visual morphology, M$_{\rm stellar}$ $>$ 5$\times$10$^{10}$ M$_{\odot}$ and z$_{spec}$ $\lesssim$ 1
(to avoid severe cosmological dimming effects). We find 6 objects satisfying these criteria. 
These galaxies are also the most massive within the HUDF up to this redshift limit. 
Our galaxy sample was firstly identified by means of the Rainbow database\footnote{https://rainbowx.fis.ucm.es/} \citep{Perez-Gonzalez08a,Barro11a,Barro11b}.
 %Rainbow consists of a vast compilation of photometric and spectroscopic data
 %for several premium fields (HUDF, GOODS North and South, EGS, COSMOS, etc.)
 %accessible via an online browser tool which provides the user with many
 %capabilities such as building Spectral Energy Distributions (SEDs) covering from
 %X-rays to radio wavelengths. 
 
Spectroscopic redshifts are available for our whole sample \citep{Croom01,Vanzella05,LeFevre05,Ravikumar07}.
In order to be self consistent and to use the information in the HUDF images, instead of using the Rainbow mass estimates,
we performed SED fitting using the Le Phare photometric redshift code \citep{Arnouts1999,Ilbert06}
to obtain stellar masses for each object based on the the total fluxes derived from the 4 S\'ersic component fits plus residuals (see Section \ref{subsec:psf}).
A range of short duration tau models (30, 70, 100, 300 Myr e-folding time) and a burst model
were included in the template set.  The models were produced using \citet{Bruzual03} (BC03)  at solar metallicity
with a \citet{Chabrier03} IMF. The fitted ages were required to be younger than the age of the Universe
at the redshift of the source, and no dust extinction was allowed in the fitting,
because it is expected to be of negligible importance for massive ETGs. The outcomes
for our galaxy sample are listed in Table \ref{tab:galaxies}. 

We also supplement the table 
with the masses changed to a \citet{Salpeter1955} IMF (+0.23 dex, as in \citet{Cimatti08}) due to increasing evidence for a more bottom-heavy IMF for massive galaxies
\citep{LaBarbera13,Ferre-Mateu13,Martin-Navarro15}. 
We stress that, according masses derived with a Chabrier IMF, HUDF-1 falls below our mass cut.  
However, given that the mass derived with a Salpeter IMF does meet our criteria, we chose to 
keep this object in our sample as it is among the most massive
objects in HUDF at $z <$ 1.

A montage with the galaxies in our sample is shown in Figure \ref{fig:mosaicluminanceIR}.
The ubiquity of morphological low surface brightness features displayed by these galaxies is noteworthy (like the shells in HUDF-3 or the fan of stars in HUDF-5).
In addition, a large number of minor objects surrounding the massive galaxies are present. 
It is beyond the scope this paper to identify them as galactic satellites, but we would expect to see a large number of satellites if minor merging is significantly contributing to the evolution of massive galaxies \citep{Bluck12,Newman12,Lopez-Sanjuan12,Marmol-Queralto12,Marmol-Queralto13,Ferreras14,Ruiz15}.

\begin{center}
\begin{table*}
\caption{List of filters}
\label{tab:filters}
\begin{tabular}{cccccc}
\hline
Instrument & Filter & Exposure time & Zeropoints & PSF FWHM & Pixel scale\\
                  &          & [sec]               &  [mag]         & [arcsec]    & [arcsec/pix]\\
\hline
ACS    & F435W & 134880 & 25.673 & 0.080 & 0.03\\
ACS    & F606W & 135320 & 26.486 & 0.073 & 0.03\\
ACS    & F775W & 347110 & 25.654 & 0.080 & 0.03\\
ACS    & F850LP & 346620 & 24.862 & 0.085& 0.03\\
WFC3 & F105W & 333877 & 26.269 & 0.181 & 0.06\\
WFC3 & F125W & 193307 & 26.230 & 0.185 & 0.06\\
WFC3 & F140W & 82676 & 26.452 & 0.187 & 0.06\\
WFC3 & F160W & 317944 & 25.946 & 0.190 & 0.06\\
\hline
\end{tabular}
\end{table*}
\end{center}

\begin{center}
\begin{table*}
\caption{List of galaxies}
\label{tab:galaxies}
%tables created automatically in the file for_table.txt
\begin{tabular}{ccccccccccc}
\hline
Galaxy name & R.A.                & Dec.     & z$_{spec}$    & Mass$_{\rm Chabrier}$ &  Mass$_{\rm Salpeter}$    & r$_{e,\rm H-band}$ & r$_{e,\rm circ,H-band}$ & axis ratio & Pos. angle\\
                     & [J2000]           & [J2000] &                     & [log($M_{\odot}$)]        &  [log($M_{\odot}$)]           & [arcsec]       & [kpc]               &   b/a         & [degrees]           \\
\hline
HUDF-1  & 53.16161 & -27.78030  & 0.619  & 10.42$^{+0.03}_{-0.03}$ & 10.65 & 0.34$\pm$0.02              & 1.70$\pm$0.15                   & 0.54$\pm$0.01          & 33.44$\pm$0.10          \\
HUDF-2  & 53.17253 & -27.78817  & 0.622  & 10.81$^{+0.16}_{-0.03}$ & 11.04 & 0.63$\pm$0.06              & 3.06$\pm$0.35                   & 0.52$\pm$0.01          & -47.07$\pm$0.03           \\
HUDF-3  & 53.14893 & -27.79976  & 0.664  & 10.90$^{+0.05}_{-0.01}$ & 11.13 & 0.42$\pm$0.03              & 2.66$\pm$0.21                   & 0.81$\pm$0.01          & -26.77$\pm$0.10          \\
HUDF-4  & 53.16341 & -27.79962  & 0.665  & 10.81$^{+0.07}_{-0.03}$ & 11.04 & 0.25$\pm$0.02              & 1.59$\pm$0.11                   & 0.83$\pm$0.01          & 22.04$\pm$0.08          \\
HUDF-5  & 53.15543 & -27.79156  & 0.667  & 11.19$^{+0.09}_{-0.05}$ & 11.42 & 0.63$\pm$0.05              & 4.16$\pm$0.34                   & 0.90$\pm$0.01          & 75.18$\pm$0.07          \\
HUDF-6  & 53.15491 & -27.76895  & 1.096  & 11.43$^{+0.00}_{-0.03}$ & 11.66 & 0.68$\pm$0.05              & 4.54$\pm$0.32                  & 0.68$\pm$0.01          & 68.52$\pm$0.04          \\
\hline
\end{tabular}
\end{table*}
\end{center}

\section{The analysis}
\label{sec:analysis}

The survey images were carefully reduced and sky-subtracted \citep{Koekemoer13}. We created 400 kpc wide postage stamps 
to explore the light distribution around the galaxies in the 8 filters available.
We masked the neighbouring objects using SExtractor-based \citep{Bertin1996} optical and NIR masks, 
which were later visually inspected and modified to remove any spurious light contribution. 
The final depictions of our masks upon the galaxy images are shown in Figures \ref{fig:masks_nir} and \ref{fig:masks_opt}.
The displayed images are the coaddition of all NIR and optical bands and they have also been smoothed by
a 2-arcsec standard deviation Gaussian kernel. These choices have been taken in order to highlight the lowest signal-to-noise features in the images.
It is also fair to say that, despite the generous masking that has been applied, it is impossible to get rid of every single source of contamination.
Nevertheless, after this careful effort, all the massive galaxies steadily decrease their surface brightness profiles down to the detection limits (golden ellipses).

We also require very accurate local sky subtraction as any residual background hampers our
 efforts for exploiting the extraordinary depth of our imaging. 
This aspect is particularly relevant if one is to sample
very faint surface brightness features, and we proceeded as in \citet{Trujillo13}.
We determined that the sky noise was dominant at galactocentric distances higher than 120 kpc for all galaxies.
Therefore, we estimated the sky level in each image at a radial distance of 140 $<$ R $<$ 160 kpc and subtracted that value.
This meticulous analysis enables us to detect galaxy light down to 31 mag arcsec$^{-2}$ (3$\sigma$ in 10$\times$10 arcsec boxes),
consistent with the limiting magnitude determinations in \citet{McLure13b}.
%This meticulous analysis enables us to detect galaxy light down to a limit (31 mag arcsec$^{-2}$ for all images) consistent with the
%HUDF12 reported depth in every filter. \textbf{This surface brightness threshold comes from the determinations in \citet{McLure13b} where
%they give 29.5-30 mag for 5 sigma detections in $\sim$0.5 arcsec apertures in contrast to our values based on 2 arcsec apertures and going down to 3 sigma detections.}

For sampling the galaxy surface brightness profiles from our galaxy sample, we created concentric
elliptical apertures from the galaxy centre, 0.5 kpc wide in the inner 2 kpc, and 2 kpc wide at greater distances. We fixed the axis ratio and position angle 
of these elliptical apertures to the H-band single S\'ersic outputs (see Subsection \ref{subsec:psf}), in order to
sample consistently the surface brightness profiles for all filters. The H-band filter is chosen because it is the reddest and as such it is the most representative of the total stellar component.
In those annuli we estimated the galaxy flux by the 3$\sigma$ clipped mean
of the pixel values on those apertures, and then we apply the formula

$$\Sigma \mathrm{[mag/arcsec^{2}]}=-2.5 \mathrm{log}(F_{annulus})+zp+5 \mathrm{log}(S_{pix})$$

where $\Sigma$ is the galaxy's surface brightness, $F_{annulus}$ is the average flux per pixel within the annulus, $zp$ stands for each image zeropoint and $S_{pix}$ is the pixel scale 
(0.06 arcsec/pix for WFC3 and 0.03 arcsec/pix for ACS). The only object that was not totally explored using this method
is HUDF-3, whose WFC3 images do not cover the whole galaxy (see Fig. \ref{fig:mosaicluminanceIR}).

\begin{figure*}
% \begin{center} 
\vspace{0.5cm}
\hspace{2cm}
\rotatebox{0}{
\includegraphics[angle=0,width=1.0\linewidth]{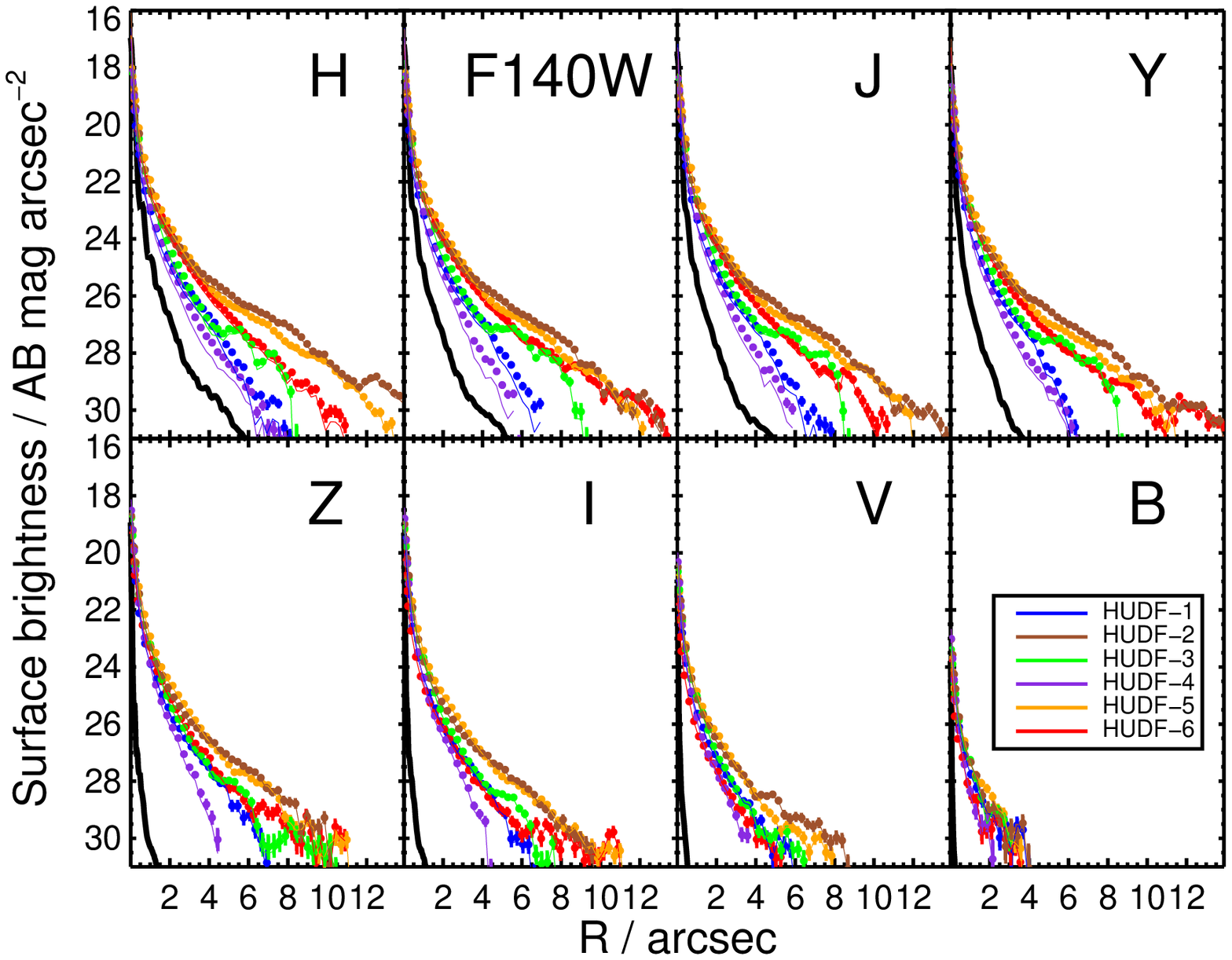}}
% \end{center}
\vspace{-0.5cm}

\caption{Comparison among the observed galaxy surface brightness profiles (coloured points), the best galaxy surface brightness models (``a la Szomoru": deconvolved profiles + residuals; coloured lines) and the PSF profiles (scaled up to match the galaxy centres; black lines) for each HST band. As expected for large early-type galaxies in space observations (small PSF FWHM) the convolved and unconvolved profiles are not very different. Moreover, the outer parts of the galaxies do not decrease their brightness in a similar way than the PSF profiles, limiting any ``red halo" issue in our sample.}
\label{fig:psf_compared_to_gals}

\end{figure*}

\begin{figure*}
% \begin{center} 
\vspace{0.5cm}
\hspace{2cm}
\rotatebox{0}{
\includegraphics[angle=0,width=1.0\linewidth]{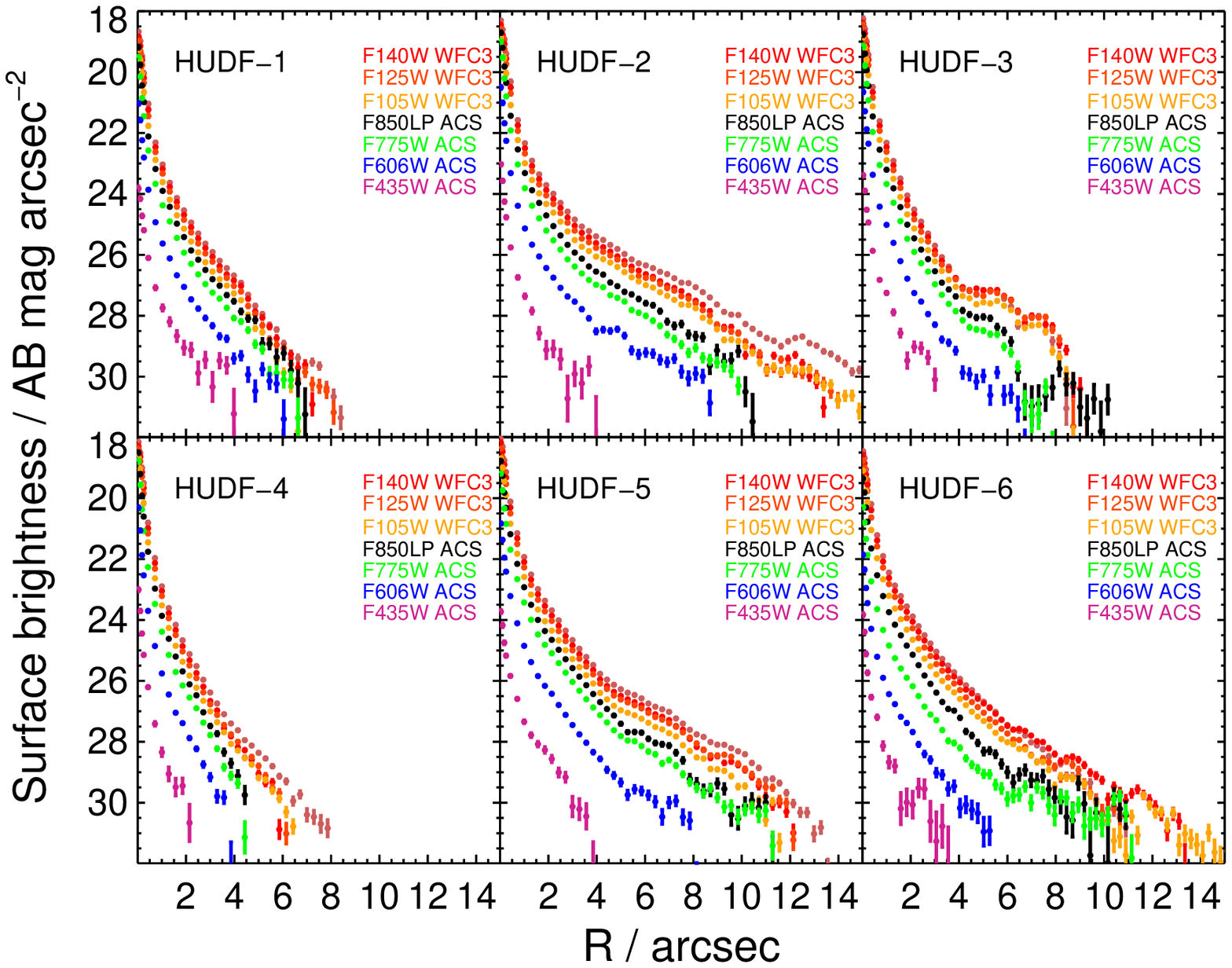}}
% \end{center}
\vspace{-0.5cm}

\caption{Observed surface brightness profiles measured within each of the HST filters available
for our ETG sample. Each individual point was calculated in elliptical 2 kpc
wide apertures (except for the central four points where 0.5 kpc wide apertures
were used), applying a 3$\sigma$ clipped mean in those annuli, for retrieving
the surface brightness values and the associated error bars. For all cases, these massive ETGs are
more luminous and extended in the redder bands. The galactocentric distances
probed in this study, sometimes more than 100 kpc at z = 0.6 - 1, are comparable
with local Universe ETG very deep observations \citep{Kormendy09,Tal11}.}
\label{fig:mosaicprofiles}

\end{figure*}

\subsection{Surface brightness fitting and the impact of the PSF}
\label{subsec:psf}

The Point Spread Function (PSF) of the images not only sets the angular resolution of our observations but also determines how
the galaxy light is scattered \citep[see for a recent analysis][]{Sandin14,Sandin15}. Hence, correcting the observed surface brightness profiles by the PSF distortion is 
essential to retrieve accurate 2D surface brightness maps and structural parameters. 
To that end, we have fitted using GALFIT \citep{Peng10},
from 1 to 4 S\'ersic functions to all the images of the galaxies within our sample. The reason behind
our multicomponent fitting is to ensure that we are describing
the 2D distribution of each galaxy's light to the greatest level of detail
permitted by our privileged photometry, avoiding any possible overmodelling ($\chi^{2}_{\nu}<$1). By so doing,
it is important to realize that we cannot give any physical interpretation to the different S\'ersic function
fits to the galaxy surface brightness profiles in ETGs, as done by other studies focused on late-type galaxies \citep{Zibetti04b,Trujillo13},
without the addition of kinematic information \citep{Falcon-Barroso06,Krajnovic08,Krajnovic13}.

The S\'ersic functions are axisymmetric and as such, it is impossible (unless
you perform an ad-hoc fit in a particular set of pixels of your image) to model
any non symmetric substructure in the galaxy's surface brightness profiles.
We thus selected as the best galaxy model the 4-S\'ersic deconvolution adding the residuals
of the fit --as done in \citet{Szomoru12}, hereafter ``a la Szomoru'' method-- trying to capture any possible feature not represented by the symmetric S\'ersic
functions. Contrary to this ``a la Szomoru'' method, we masked the central 10 pixels when performing the residual addition 
 as these central pixels have some artificial noise owing to the exact positioning of the PSF peak.
Please see the Appendix \ref{appendix:fits} for a comparison of these ``a la Szomoru” surface brightness 
profiles with the rest of the fits, as well as the observational galaxy profiles.

It is to note that neighbour galaxies have been masked but not subtracted in our GALFIT fits, and thus a certain level of light contamination is expected.
We ensure that in all cases (except HUDF-6), none of the 5 brightest objects in each galaxy stamp is within our limit for the surface brightness
determination. For our exception, HUDF-6 images show a companion star at $\sim$10 arcsec from the galaxy's centre. The difference
between fitting or not (using of course a PSF model) this star for the galaxy total flux is $<$ 0.02$\%$, and thus negligible.

Our PSF choice must not only be accurate but very extended as well, in order to prevent any red spurious excess at large
radii mimicking the light contribution of a stellar halo \citep[the so-called "red halo" problem, e.g.][]{Zibetti04a, Zibetti04b, Zackrisson06, deJong08}.
In theory, we should go as far as 1.5 times the full galaxy size \citep{Sandin14,Sandin15,Trujillo16}.
Tiny Tim \citep{Krist1995} is the only way to build such extended HST PSFs.
Therefore, we created our Tiny Tim simulated stars by assuming they should extend up to the equivalent size of 200$\times$200 kpc
at the median redshift ($<z>=$ 0.65) of our galaxy sample. This translates into PSF sizes of 500$\times$500 pixels
for WFC3 and 1000$\times$1000 pixels for ACS. However, for ACS images, 
Tiny Tim cannot retrieve models spanning such large distances, and thus we content ourselves with the maximum extent possible for this camera. However, this fact has very little (if any)
impact in our analysis because of the very small sizes of our passive galaxy sample in the bluest bands.

We further improved the PSF produced by Tiny Tim in each band by replacing the core with that of an isolated non-saturated star
at RA=03:32:38.01, DEC=-27:47:41.67 (J2000). This mitigates the effect shown by Bruce et al. (2012) whereby Tiny Tim underpredicts
the PSF flux at distances greater than 0.5 arcsec. We also rotated these hybrid stars in order to match the position of the stellar spikes in HUDF science image.
The chosen star’s spectral type is K4-K5 star \citep{Pirzkal05}, which is optimal for studying early-type galaxies as the light from both the star and the galaxies is scattered
similarly in broadband filters \citep{LaBarbera12}.

After these considerations on our PSF model, we checked about the existence of the ``red halo" problem in our sample. The test we performed is to be found in Figure \ref{fig:psf_compared_to_gals}. We compare there our observed surface brightness profiles, the ``a la Szomoru" profiles and the PSF profiles (scaling them up to match the peak in the galaxies' surface brightness profiles). These are space observations (small PSF FWHM), and as such there is not so much difference between the convolved and unconvolved + residuals profiles. It is also reassuring that the outer parts of the galaxies do not decrease in a similar way as the PSF profiles, and thus limiting the impact of the ``red halo" problem. What we find is that there is an exponential decay for HUDF-2 in the NIR bands between 3 and 8 arcsec, indicating the presence of an inner disk.
We will further discuss about it in Section \ref{subsec:ongoing}. Finally, we did another similar exercise comparing star ang galaxy profiles, but normalizing this time the PSFs to the total galaxy fluxes. We obtained similar conclusions.

\section{Results}
\label{sec:results}

We show our observed surface brightness profiles in Figure \ref{fig:mosaicprofiles}. It
 is worth noting that the various galaxies in our sample show emission extending to different
 galactocentric distances and that none of them have signs of abrupt truncation even
at the faint levels explored. Every galaxy is more extended and more luminous in the redder bands
 as expected for passive ETGs. For some of the objects, we reach 10-12 arcsec
 in the H-band, which is comparable to local Universe
 very deep observations \citep{Kormendy09,Tal11} but this time at a median redshift $<z> =$ 0.65
where the cosmological dimming make all galactic features $\sim$2 mag arcsec$^{-2}$ fainter.

\begin{figure*}
% \begin{center} 
\vspace{0.5cm}
\hspace{2cm}
\rotatebox{0}{
\includegraphics[angle=0,width=1.0\linewidth]{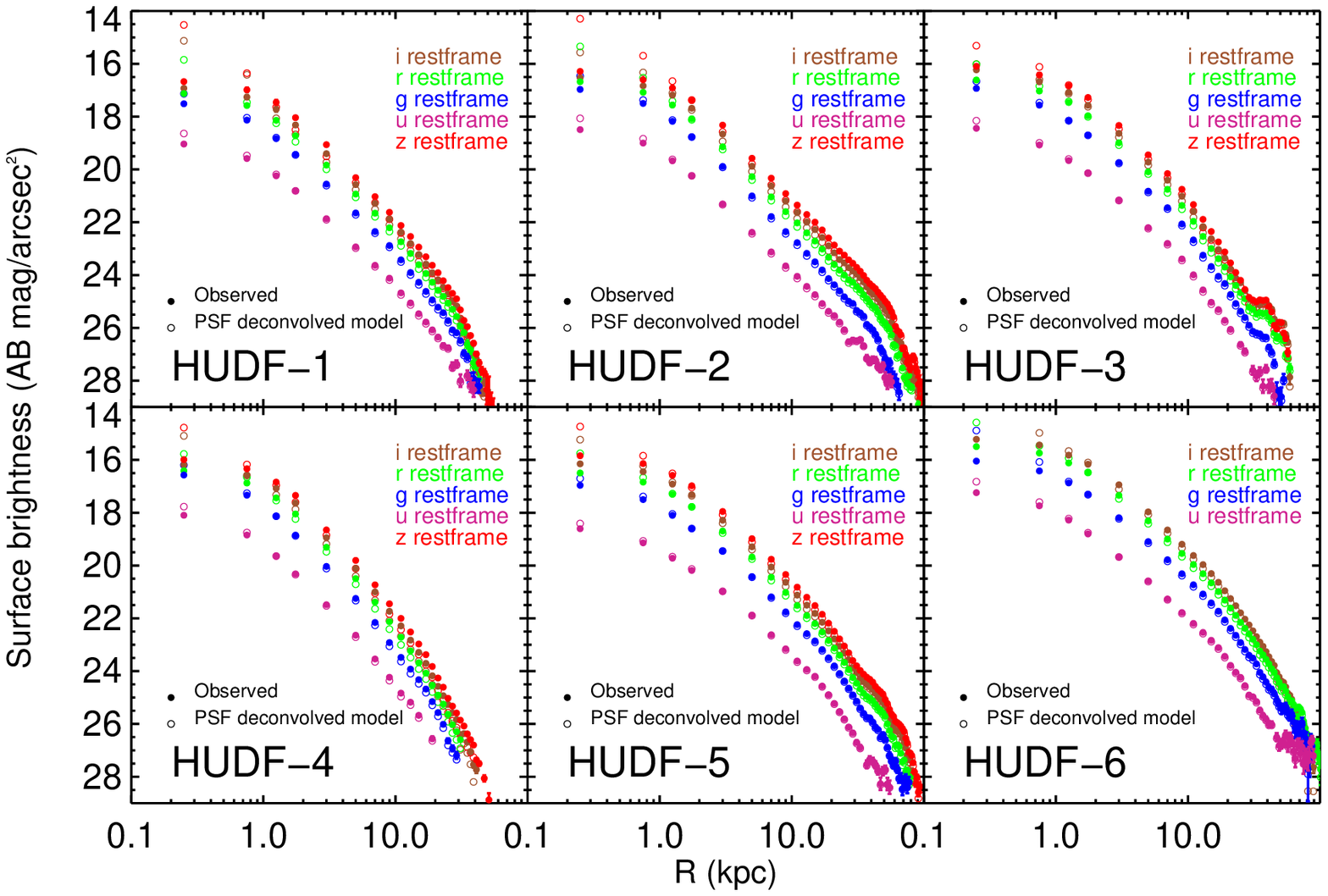}}
% \end{center}
\vspace{-0.5cm}

\caption{The $u$, $g$, $r$, $i$ and $z$-band Sloan filters equivalent restframe surface brightness
profiles for the six galaxies in our sample. They were created by linearly interpolating
the HST filters, both for the observed and the model$+$residual ``a la Szomoru" profiles, and then 
correcting the surface brightness by cosmological dimming. Note that, for HUDF-6, z-band is not covered
due to its redshift ($z\sim$1.1). It is clear that the PSF effect scattering the
light coming from these objects is more pronounced for the inner galaxy parts.
It is also interesting checking that HUDF-2, HUDF-3 and HUDF-5
have bumps at restframe surface brightness 25-26 mag arcsec$^{-2}$, and they are specially strong in the redder
bands. By joining this information with their visual appearance, we associate these
features to recent merging events.}
\label{fig:restframe_profiles}

\end{figure*}

\begin{figure*}
% \begin{center} 
\vspace{0.5cm}
\hspace{2cm}
\rotatebox{0}{
\includegraphics[angle=0,width=1.0\linewidth]{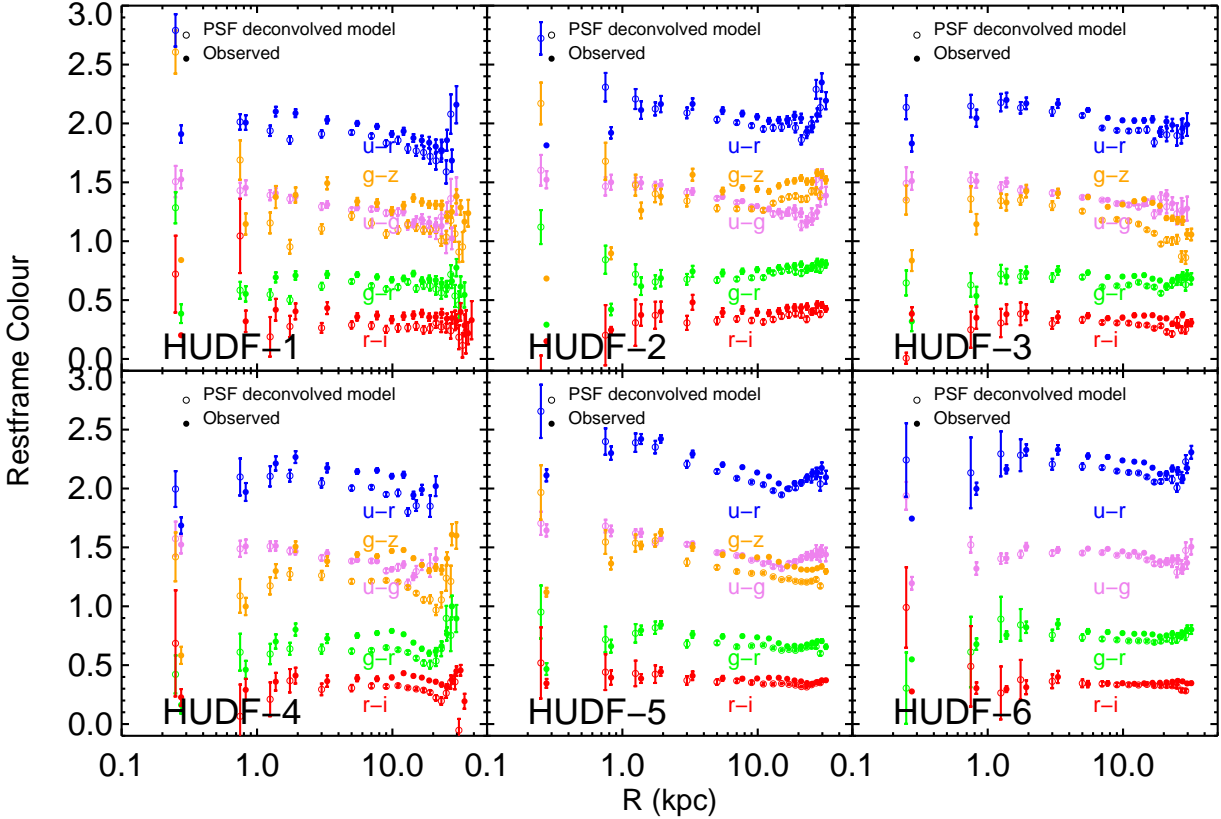}}
% \end{center}
\vspace{-0.5cm}

\caption{The $u-g$, $u-r$, $g-r$, $g-z$ and $r-i$ Sloan filters equivalent restframe colour profiles for the
six galaxies in our sample. Both observational and model$+$residual ``a la Szomoru" profiles area
plotted (with a slight shift in the x-axis for a better reading), along with their errors up to the limit of 30 kpc.}
\label{fig:restframe_colors}

\end{figure*}

%\begin{figure*}
%% \begin{center} 
%\vspace{0.5cm}
%\hspace{2cm}
%\rotatebox{0}{
%\includegraphics[angle=270,width=1.0\linewidth]{test_new_distances.ps}}
%% \end{center}
%\vspace{-0.5cm}
%
%\caption{We show the colors with a (base 2) logaritmic sampling in galactocentric distance 
%in order to reduce the errors in the galaxy outer parts, adding $g-z$ and $u-r$ to our previous results.
%The trend for redder colors in the galaxies undergoing merger episodes is still preserved.}
%\label{fig:restframe_colors_new_distances}
%
%\end{figure*}

\begin{figure*}
% \begin{center} 
\vspace{0.5cm}
\hspace{2cm}
\rotatebox{0}{
\includegraphics[angle=90,width=1.0\linewidth]{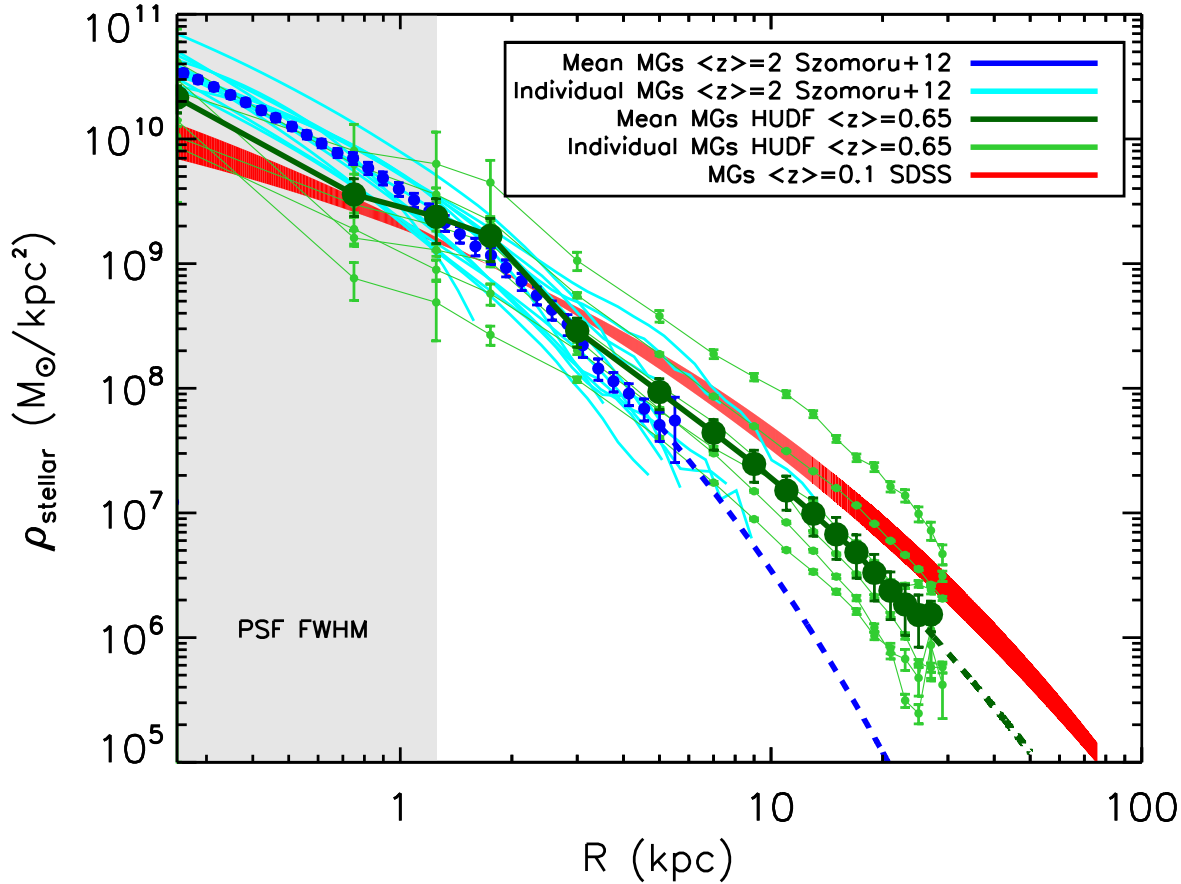}}
% \end{center}
\vspace{-0.5cm}

\caption{The circularized stellar mass density profiles for the Massive Galaxies (MGs) in our sample, comparing
them with similar mass SDSS ETGs and the massive compact galaxies in \citet{Szomoru12}.
Individual mass profiles are shown in light colours, while the mean profiles are in dark colours
(and their extrapolations are the dashed lines).
HUDF massive galaxies show an excess of mass in their outer parts, opposite to what could be
seen for the high-z sample, and closer what was found for local massive ETGs. This evidence points
to the progressive building up of stellar haloes as the link between the two other populations.}
\label{fig:circularized_mass_profiles}

\end{figure*}

\begin{figure*}
% \begin{center} 
\vspace{0.5cm}
\hspace{2cm}
\rotatebox{0}{
\includegraphics[angle=0,width=1.0\linewidth]{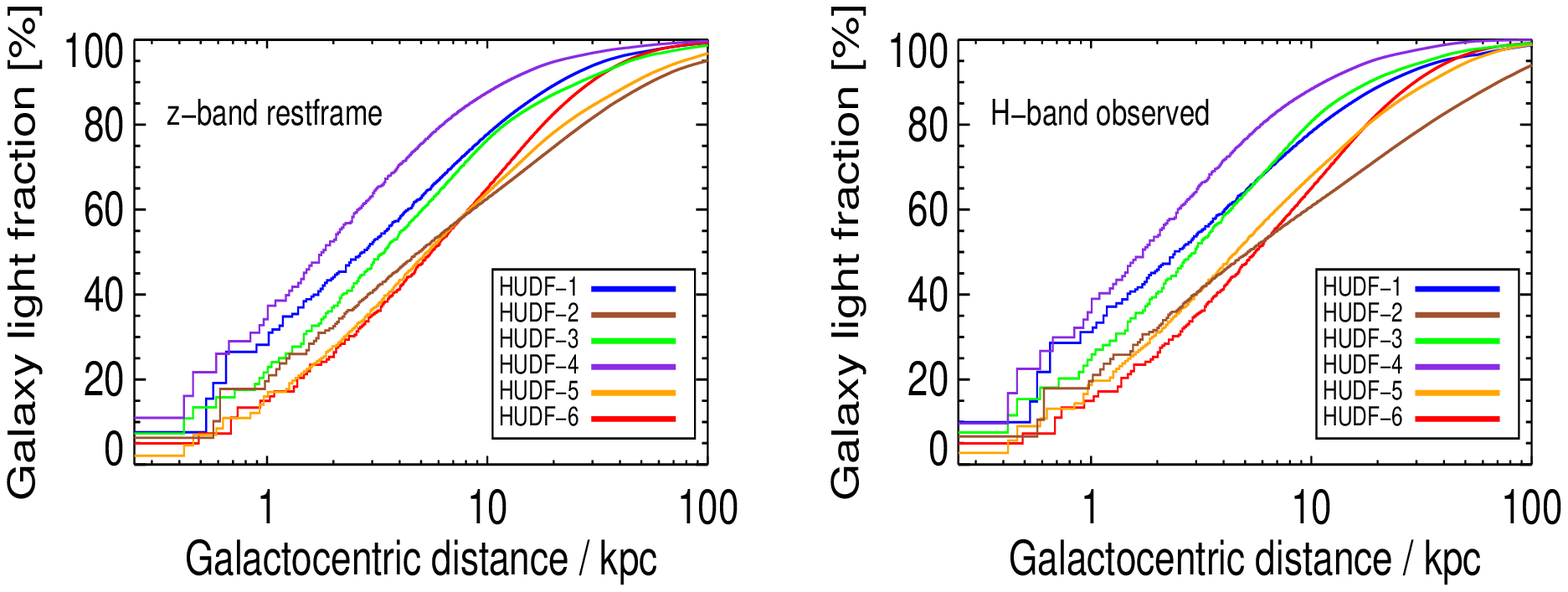}}
% \end{center}
\vspace{-0.5cm}

\caption{Cumulative light fractions for our inferred z-band restframe (the base for our mass profiles) and the reddest observed band (H).
Except for HUDF-4, the most compact galaxy, the rest of the massive galaxies store between 20 and 40\% of their light at galactocentric
distances greater than 10 kpc.}
\label{fig:cumulatives}

\end{figure*}

\subsection{Sloan equivalent filters and colours}
\label{subsec:sloan}

We have calculated Sloan bands equivalent restframe surface brightness
profiles for the six galaxies in our sample (Fig. \ref{fig:restframe_profiles})
for determining colours and masses at each step in galactocentric distance. They were constructed from both the observed
and the model$+$residual ``a la Szomoru" profiles by linearly interpolating
the HST filters and then correcting the surface brightness by cosmological dimming \citep[as done before in][]{Trujillo13}.
It is noticeable that the PSF effects are more pronounced for the central parts where the galaxy flux is more concentrated, and for the redder filters,
as the WFC3 PSF is broader than the ACS one. As expected, correcting
for the PSF produces brighter galaxy cores and slightly fainter profiles at intermediate galactocentric distances, while at larger distances
($>$ 30 kpc) the effect is almost negligible. For the galaxies HUDF-2, HUDF-3 and HUDF-5,
a number of quite distinctive surface brightness bumps at magnitude $\sim$25 are visible. They are especially
strong for the redder bands. In the latter two cases, the association with recent merger
events is evident, looking at the visual morphologies in the NIR bands.
For the remaining one, this may be also the case, as it looks very
asymmetric in the same photometric bands. 

With these profiles, we computed the Sloan filters equivalent $u-g$, $u-r$, $g-r$, $g-z$ and $r-i$ colours in Figure \ref{fig:restframe_colors}
up to 30 kpc. The inner parts of the colour profiles are uncertain due to the few pixels that enter in our concentric ellipses for the surface brightness
calculations. For instance, observing with WFC3 (0.06 arcsec/pix) a galaxy at $z$ = 0.65 ($\sim$7 pix/arcsec) means that the inner kpc 
is comprised in a radius of $\sim$ 2 pixels. After the central kpc, the profiles are rather flat. We choose not to show beyond 30 kpc because of the
larger error bars ($>$ 0.2 mag) and also upbends in some profiles. The reason behind these odd colours at large galactocentric distances
is the aggressive sky subtraction, as the HUDF was optimized to look for high-z galaxies, and therefore any very extended structure is
affected even with our careful data reduction.

%These redder colors at large radii could come either from an
%aggressive sky subtraction or could be the signature of merging with minor objects
%that are older and more metal poor than the galaxy colore \citep{LaBarbera12}. Our choice is to take a conservative approach,
%limiting our analysis to the point (30 kpc) were our color profiles are robustly inferred. The only exception is HUDF-4,
%the smallest galaxy in our sample, where for the ``blue" color $u-r$ only reaches 20 kpc.

%We reduced the errors bars in Fig. \ref{fig:restframe_colors_new_distances}
%by increasing our spatial binning with a (base 2) logarithmic scale in galactocentric distance.
%We also supplement the $g-r$ and $u-r$ colors that will be later utilized for deriving the mass density
%profiles. The trends for redder colors in the external parts are still present as in Figure \ref{fig:restframe_colors}, reassuring us in our interpretation.
%although for HUDF-6 the outskirts appear to be bluer. This galaxy is the
%highest redshift ($z_{\rm spec} =$ 1.096) object in this work, and thus its outer stars are younger than for the rest of massive galaxies in our sample. 

\subsection{Stellar mass profiles}
\label{subsec:mass_profiles}

Figure \ref{fig:circularized_mass_profiles} 
shows the circularized stellar mass density profiles for the galaxies in our sample.
We calculated them using the prescriptions in \citet{Roediger15} Appendix A Table 2 for mass-to-light ratios using Sloan colours. Specifically, for the galaxies at $<z>=$ 0.65, our choice of colour and
base profile was $g-z$ and $z$, and for HUDF-6 we utilized $u-r$ and $r$. These sets of colours and bands
were chosen in order that the blue band is only constructed from the ACS filters and the red band uses only WFC3 information.
Consequently, we avoid any PSF mismatching effects that may arise
in case one combines photometric data coming from two different cameras.
We normalized the total masses by those obtained from the SED fitting.
Overplotted are the mass profiles for similar mass
(8$\times$10$^{10}$ $<$ M$_{\rm stellar}$ / M$_{\odot}$ $<$ 1.2$\times$10$^{11}$) ETG galaxies (S\'ersic index $n >$ 2.5) in NYU catalog \citep{Blanton05}
at 0.08 $< z <$ 0.12 (the uncertainties are given as a shaded red region) and the 
massive and compact galaxies in \citet{Szomoru12} at 1.75 $< z <$ 2.5 (with mean $r_{\rm e,circ}$ = 0.98 kpc and $n =$ 3.92). 
For both our sample and Szomoru et al.'s, we provide the individual and mean profiles,
We choose to stop ours at 30 kpc in order not to be affected by
any colour uncertainties in our light-to-mass conversions. Our sample of massive HUDF ETGs show extended
stellar haloes not present in the compact high-z galaxies \citep{Bezanson09,Cassata10,Szomoru12,Trujillo14}, thus showing closer resemblance to the SDSS local counterparts.
%although with a mass excess in the center which will be likely relaxed by secular processes over time.

In order to parametrize this variation, we have integrated these mass mean profiles between 10 and 50 kpc,
where we can compare our results with state-of-the-art simulations \citep[][see Section \ref{subsec:sims} in the present paper]{Cooper13}.
As neither Szomoru's nor our mass profiles extent to that distance, we fit the mean profiles in the two cases to S\'ersic functions and extrapolate these functions
up to 50 kpc. Then we integrate these functions between 10 and 50 kpc.
The results are remarkable: while 3.5\% of the galaxy mass is enclosed at these distances for Szomoru et al.'s case ($<z> =$ 2),
the fraction is 15.1\% at $<z> =$ 0.65 and 28.7\% at $<z> =$ 0.1. Despite the fact that the total stellar mass for the
three mean profiles is similar ($\sim$10$^{11}$ M$_{\odot}$), the mass profiles of massive ETGs at high-z
are intrinsically different than those at lower redshifts.

In figure \ref{fig:cumulatives} we provide a more in-depth quantification of the amount of light
(both for the reddest filter, the H-band, and the z-band restframe which is the band we used to build the mass profiles) contained in the galaxies of our sample using the same elliptical apertures 
we utilized for deriving the surface brightness profiles. Between 20\% and 40\% of the light 
is distributed at distances beyond 10 kpc. 
The only massive galaxy that differs slightly (more light concentrated
in the inner parts and less in the outskirts) is the compact HUDF-4.
It is not possible to discern any sharp transition between the galaxies' cores and their external parts by either visually inspecting these plots or the mass profiles in Fig. \ref{fig:circularized_mass_profiles}.

\subsection{Comparison with state-of-the-art simulations}
\label{subsec:sims}

In this subsection we compare our observational results with the theoretical
models of \citet[][hereafter C13]{Cooper13}. These simulations
use a semi-analytic model of galaxy formation \citep{Guo11} in combination
with a cosmological N-body simulation \citep{Boylan-Kolchin09} to predict
the surface mass density profiles of  $\sim$1900 galaxies hosted by dark matter
haloes of mass 10$^{12}$-10$^{14}$ M$_{\odot}$.

In simulations it is possible to distinguish stars that are
accreted by galaxies from so-called in-situ stars formed directly in their host dark matter haloes. In observations,
the various subcomponents of late-type galaxies follow different light distributions, allowing
the canonical bulge-disk-halo decomposition \citep[e.g.][]{Trujillo13}.
In ETGs, however, both in situ and accreted
stars are distributed in spheroidal components that cannot be separated
unambiguously by decomposition of their surface brightness profiles. To
proceed, we make use of the fact that the C13 models predict that accreted
stars have much lower binding energies on average than in situ stars, with
the result that essentially all stellar mass beyond a certain
galactocentric radius
is accreted. The mass obtained by integrating both observed and
simulated mass profiles outwards from a sufficiently large radius therefore
provides a fair point of comparison, even though
it does not correspond to the total mass of accreted stars in either case.

%As stated in Section \ref{subsec:psf},
%the various subcomponents of late-type galaxies follow different light distributions, allowing
%the canonical bulge-disk-halo decomposition \citep[e.g.][]{Trujillo13}. This is, of course, not possible
%for ETGs and therefore it is very complicated to determine the importance for ETG stellar haloes because there is not
%an unambiguous definition for this galaxy component. However, we can take advantage of the fact that stellar haloes are composed of accreted stars as opposed to 
%those formed in-situ, and that we can track both in simulations. Thinking of a pure inside-out growth scenario we could assume that most of the mass 
%in the external parts (haloes) of our sample of galaxies is accreted mass. Then, by integrating our observational mass profiles and
%simulated accreted stars from a radius far from the galaxy center, we are comparing how much mass
%is enclosed in ETG stellar haloes in both observations and simulations.

In the C13 simulations, late and early types are separated by the ratio of
bulge to total mass predicted by the \citet{Guo11} model (B/T less or
greater than 0.9 respectively). The particle tagging method used by C13 to
predict surface brightness profiles introduces an additional free parameter
beyond those of the Guo et al. model, f$_{mb}$. This controls the depth in the host
dark matter potential at which newly-formed `stars' are inserted into
the simulation.
For example, a value of f$_{mb}$=1\% means that newly formed stellar
populations initially have
a binding energy distribution identical to that of the most
tightly-bound 1\% of the dark matter in their host dark matter halo
(see C13 for details).
C13 explore a range of values for this parameter, which they find to be
strongly constrained to a range 1-5\% by the observed size-mass relation of
galaxies dominated by in situ stars (i.e. discs) at $z =$0. In
practice, the precise choice of f$_{mb}$ makes only a very marginal
difference to the
results we discuss here \citep{Cooper13,Trujillo16}. We therefore
report comparisons
against the f$_{mb}$ = 1\% results of C13.

\begin{figure*}
% \begin{center} 
\vspace{0.5cm}
\hspace{2cm}
\rotatebox{0}{
\includegraphics[angle=270,width=1.0\linewidth]{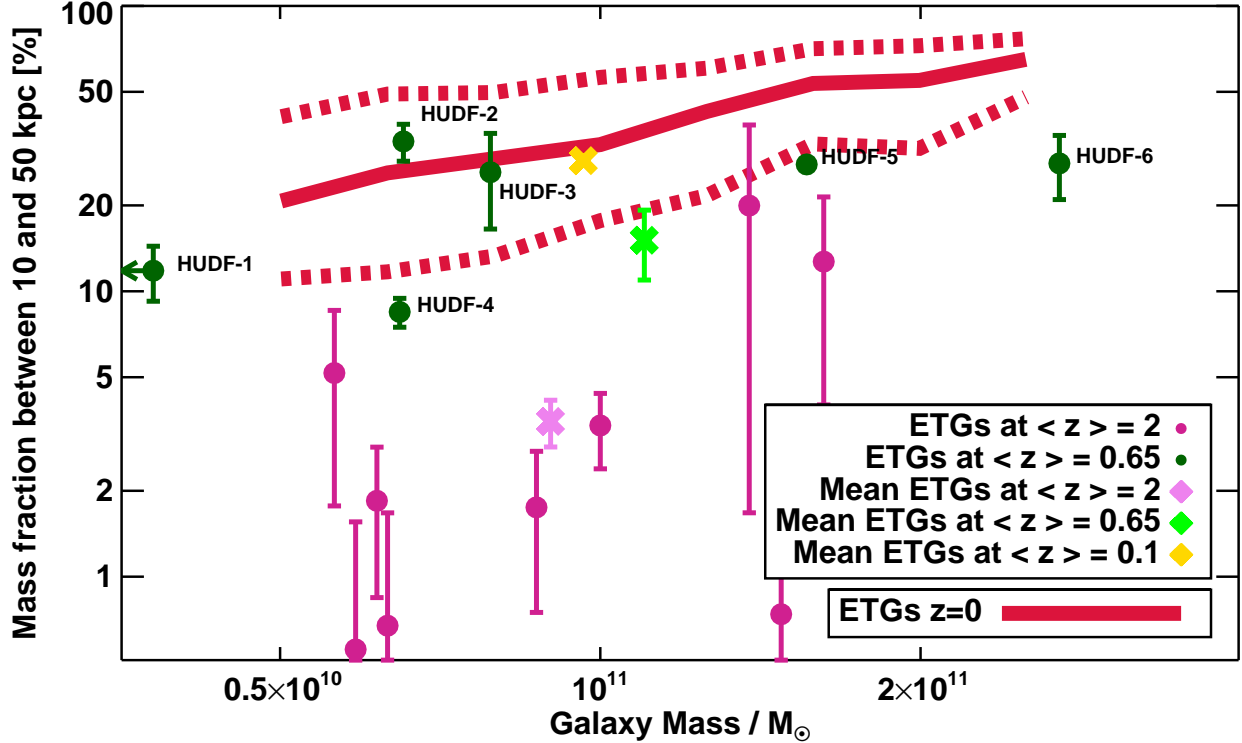}}
% \end{center}
%\vspace{-0.5cm}

\caption{Fraction of the galaxy stellar mass between 10 and 50 kpc versus the total stellar mass for our sample of six Early-Type Galaxies (ETGs; the green points).
Our inferred mass fractions come from the recipes in \citet{Roediger15}, and the errors from the absolute differences between those and the ones in \citet{Bell03}
(HUDF-5's errors are too small to be seen).
The solid red lines are the results for local ETGs in \citet{Cooper13} simulations, with the dashed lines corresponding to the 16-84 percentile range.
The violet data points are individual massive galaxies at $< z >=$ 2 studied in \citet{Szomoru12}, but we want to emphasize that
they are extrapolations as information about their mass profiles is unavailable at these galactocentric distances. The crosses in light violet, light green and golden colours denote the mean
values for the $< z >=$ 2, $< z >=$ 0.65 and $< z >=$ 0.1 massive galaxy samples respectively (cf. Fig. \ref{fig:circularized_mass_profiles}).
There is a rough correlation between galaxy mass and the percentage of mass in the outskirts for the green points,
following the simulation predictions. However, it is to note that our data do not follow the local
relationship because of the median redshift of our sample ($<z>=$ 0.65).
Most importantly, 5-20\% of ETGs stellar mass is located in their ``haloes", above the 
high-z data points and in stark contrast with recent results for late-type galaxies \citep{Trujillo13, vanDokkum14, Trujillo16}.}
\label{fig:from_10_to_50_kpc}

\end{figure*}

The further from the centre of the galaxy, the lower is the contribution by in situ material to the mass profile. Being
conservative, we will start our integration from the typical distance where high-z massive 
galaxy surface brightness profiles finish ($\sim$10 kpc, see Fig. \ref{fig:circularized_mass_profiles}) and hence identify our stellar haloes as the 
light component previously missed in shallower observations. We stop our calculations at 50 kpc, our previous integration limit.
The results for our galaxy sample are plotted in Figure \ref{fig:from_10_to_50_kpc}, and their error bars stem from the difference in the mass determinations by using either the \citet{Roediger15}
or the \citet{Bell03} recipes. We also supply the local galaxy mass fraction at 10 $<$ R / kpc $<$ 50 relationships from the \citet{Cooper13} simulations for 
ETGs. This relationship is displayed in red colour, with the 16 and 84 quartiles being the dashed lines. 
For consistency, we also overplot the extrapolations for the individual massive ETGs in \citet{Szomoru12} and the mean values for the three samples we are using
throughout this paper (namely Szomoru's, HUDF and SDSS).

There is an overall departure of our galaxy sample from the local relation, most probably due to the fact that they are not
$z =$ 0 galaxies ($<z> =$ 0.65 median redshift). In fact, the six massive galaxies in our sample straddle the high-z and low-z data. 
Combining the location of the $<z> =$ 2 data points and Fig. \ref{fig:circularized_mass_profiles}, it seems that the HUDF ETGs are advancing towards the upper part of the plot 
to reach their fiducial $z =$ 0 relation. Very interestingly, there is a correlation between the total galaxy mass and fraction of mass
in the outer parts for our six galaxies, where they approximately follow the Cooper et al.'s ETG predictions.
It is also remarkable the agreement between Cooper's simulations and the SDSS mean value.

Quantitatively, Figure 8 (left side) in \citet{Trujillo13},
Figure 4 in \citet{vanDokkum14} and Figure 13 in \citet{Trujillo16} show that the haloes of M$_{\rm stellar}$ $\sim10^{10}-10^{11}$ M$_{\odot}$
late-type galaxies constitute at most 5\% of their total light at $z$ = 0. Our small but unique
sample shows that the stellar mass in massive ETG stellar haloes is larger, of the order of 5-20\% (and yet not at $z$ = 0, but at $z\sim$ 0.65). This contrast
between galaxy types must be investigated further \citep[see for instance][]{D'Souza14}, but makes sense from a $\Lambda$CDM perspective,
where the histories of ETGs should be more merger-dominated than for disky galaxies \citep{Cole00,Croton06,Purcell07,Ruiz15}, and also because ETGs do not
have a prominent disk storing a significant fraction of the galaxy's baryons.

\subsection{Constraining the merger channel for massive galaxy growth}
\label{subsec:ongoing}

Studies about merger rates always provide an indirect way to look at the assembly history of galaxies, because of the fact that what
it is measured is the mass to be accreted as opposed to accreted mass. It is interesting to see whether this could be improved by using very
deep images to trace any signature of ongoing merging.

To this end, we have created the following exercise. We take the galaxy light which is not described by the
overall galaxy spheroid, i.e., the residuals from subtracting the single S\'ersic fit to each galaxy surface brightness profile. Converting those into mass (in an
approximate manner, given the information at hand) we can check their relative importance. The reason behind this exercise resides in the fact that 
some low surface brightness features come from galaxy interations (at least in HUDF-2, HUDF-3 and HUDF-5). These 
features are smooth and, as such, very hard to be picked up as potential close pairs. HUDF12 has the potential to
detect them at intermediate redshift, opening a new perspective in the mass assembly of massive galaxies.

Nevertheless, we would like to emphasize that this is just a toy model because of fitting a single S\'ersic function to deep 
and high resolution images of most ETGs, even in the local universe, leaves residuals which have nothing to do with merging features.
This seems to be the case for HUDF-1 and HUDF-2, as their residual images display negative and positive values close to the galaxy centre
in perpendicular directions corresponding to the symmetry axes, typical of the presence of a non-subtracted inner galaxy disk (as detected in Section \ref{subsec:psf}). Therefore,
for these two galaxies, we investigated the residuals coming from the subtraction of a double S\'ersic fit instead of a single S\'ersic fit.

\begin{figure*}
% \begin{center} 
\vspace{0.5cm}
\hspace{2cm}
\rotatebox{0}{
\includegraphics[angle=0,width=0.975\linewidth]{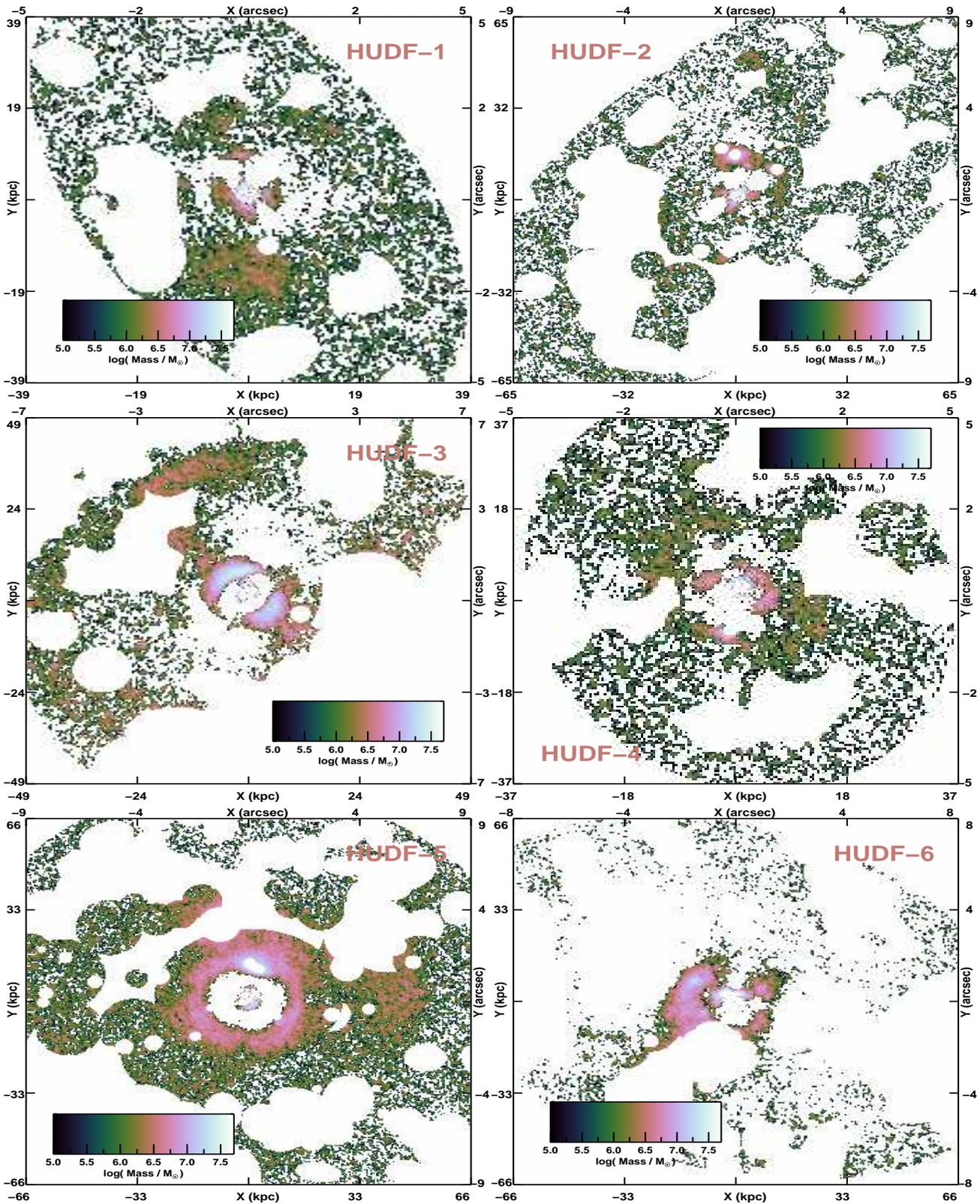}}
% \end{center}
%\vspace{-0.5cm}

\caption{Stellar mass maps corresponding to the smooth residuals in the galaxy light. Thinking of
the inside-out growth of massive galaxies, we calculated how much mass
is encompassed in minor interactions by subtracting to every galaxy a S\'ersic model of its overall
spheroid. The colour coding is the same throughout the plots, but each galaxy is shown up to its 
full extent (31 mag arcsec$^{-2}$). The white elliptical patches are the product of neighbour masking, and thus the total masses
listed in Table \ref{tab:mass_in_residuals} for these residuals (1-2\% of the total galaxy stellar mass) should be taken as a lower limit.
These numbers given by our toy model are not far from the predictions from close pairs to the mass growth of massive ETGs ($\Delta$M/M$\sim$4\% Gyr$^{-1}$).}
\label{fig:mass_in_residuals}

\end{figure*}

The step to transform from light to mass is done by a crude assumption, i.e. that the mass to light ratio is constant through the entire radial distribution. 
Considering that our galaxy sample have relatively flat colour gradients, this is reasonable. 
We utilize the value given by the MIUSCAT models\footnote{http://www.iac.es/proyecto/miles/} 
\citep{Vazdekis12,Ricciardelli12} in the reddest (SDSS i-band) mass-to-light ratio provided
assuming Kroupa universal IMF, solar metallicity and a stellar population age of 5 Gyr.
The results are given in Table \ref{tab:mass_in_residuals} and Figure \ref{fig:mass_in_residuals}. 
The errors stem from the different total amount of light enclosed in the HST residuals closest to restframe i-band.

Very little light is involved in these smooth
features (of the order of 1\% the galaxy light), with a slightly larger percentage in mass (1-2\%).
To put these numbers in context, we compare them
with best estimates from satellites/close pairs. Specifically, \citet{Ferreras14} with a sample of 238 massive galaxies at 0.3 $< z <$1.3 quantified that the upper
limit for the average mass growth rate for these galaxies is 
($\Delta$M/M)/$\Delta$t $\sim$ 0.08$\pm$0.02 Gyr$^{-1}$, while \citet{vanDokkum05} inferred
0.09$\pm$0.04 Gyr$^{-1}$ for 126 red nearby galaxies. 
To move from growth rate to mass, a timescale for the duration of the morphological
features of dry mergers should be adopted. \citet{Bell06} classified major (1:1 to
3:1) merger snapshots suggesting values of 150$\pm$50 Myr. The duration of the visibility of galaxy mergers
using CAS parameters is 0.4-1 Gyr \citep{Conselice06,Lotz08,Conselice09}. Choosing then 0.5 Gyr
as a representative number, one would expect $\sim$4\% of the total mass of the galaxy in these residuals.

Our numbers are close to these expected values, especially by thinking that some residuals in mass are not seen
because of our masking. Actually, this aspect makes our measurements a lower bound in the percentages of light and mass. Nevertheless, we believe we cannot draw any strong implications as 
this experiment has many parameters we do not control: the residuals and the mass-to-light ratios being representative of substructures,
the uncertainty about how long merging features last and cosmic variance
due to the fact of studying only six massive galaxies. It is to note that the galaxies
showing smaller residuals are two most compact ones (HUDF-1 and HUDF-4) and the most distant galaxy (HUDF-6) which
might be an indication that cosmological dimming has a deeper impact on it than for the rest of the objects, hiding some
extra mass in undetected features. Summarizing, it is very interesting to see that this naive exercise
yields numbers similar to close pairs predictions and also to check that the visually identified
merging smooth features in HUDF-2, HUDF-3 and HUDF-5 clearly provide to these galaxies with more mass in their residuals.

\begin{center}
\begin{table*}
\caption{Stellar mass contained in the residuals}
\label{tab:mass_in_residuals}
\begin{tabular}{cccc}
\hline
Galaxy & \% light in residuals & Mass                          & \% galaxy's mass   \\
       & i-band                & $\rm \times10^{8} \rm M_{\odot}$ &                 \\
\hline
HUDF-1 & 0.52$\pm$0.06 &     2.06$\pm$0.24 &     0.79$\pm$0.09 \\
HUDF-2 & 1.03$\pm$0.07 &    10.24$\pm$0.71 &     1.57$\pm$0.11 \\
HUDF-3 & 1.79$\pm$0.47 &    21.47$\pm$5.64 &     2.72$\pm$0.71 \\
HUDF-4 & 0.35$\pm$0.10 &     3.49$\pm$0.97 &     0.54$\pm$0.15 \\
HUDF-5 & 1.41$\pm$0.01 &    33.53$\pm$0.31 &     2.15$\pm$0.02 \\
HUDF-6 & 0.34$\pm$0.13 &    14.15$\pm$5.38 &     0.52$\pm$0.20 \\
\textit{Mean values}      & 0.91$\pm$0.13 &    14.16$\pm$1.69 &     1.38$\pm$0.20 \\

\hline
\end{tabular}
\end{table*}
\end{center}
 %don't forget to reshuffle the galaxies to match the new names, and to subtitute exponential annotation by powers of 10 annotation

\section{Summary and Conclusions}
\label{sec:conclusions}

We present a comprehensive characterisation of the six most massive (M$_{\rm stellar}\geq$ 5$\times$10$^{10}$ M$_{\odot}$) 
Early-Type Galaxies (ETGs) at z $\lesssim$ 1 in the deepest HST field, the HUDF. We focused our efforts in the HUDF12 programme 
\citep{Ellis13,Koekemoer13}, whose data reduction preserves extended low surface brightness features
and at redshifts where cosmological dimming is not yet strong enough ($\lesssim$ 2 mag arcsec$^{-2}$) to remove the traces of minor merging.

The substructures present in the outer parts of ETGs, whose origin is the progressive build-up of these objects via merging,
have not been studied to date at intermediate/high redshift due
to their intrinsic faintness and the very rapidly growing cosmological dimming,
which make these outskirts very challenging to detect.
Therefore, it is not yet known whether these outer parts could be described as galactic haloes, similar to those found in disk galaxies. Our work aims
to clarify this situation and investigate how massive galaxies change their observational properties since $z =$ 1. 
%A companion paper, Buitrago et al. in prep.,
%studies the implications of our study for the size-mass relation of massive galaxies.

We carefully analysed each galaxy image according to the recipes in \citet{Trujillo13}, fitting up to 4 S\'ersic functions convolved with the PSF in the 8 HST filters available.
In so doing, we are able to remove the PSF distortion in the observed profiles. Our ultradeep dataset reaches galaxy surface brightness profiles
down to 31 mag arcsec$^{-2}$ (3$\sigma$ in 10$\times$10 arcsec boxes; $\sim$29 mag arcsec$^{-2}$ after correcting by cosmological
dimming), which translates into 25 effective radii in distance, or as far as 100 kpc in some cases at an outstanding median redshift of $< z >= $0.65.

The striking difference between previous shallower observations and the HUDF12 
is the appearance of extended low surface brightness envelopes (or stellar haloes) for each individual galaxy.
Even though the small statistical representativeness of our sample, containing only 6 objects,
%Previously, \citet{Tal11} stacked 42000 Luminous Red Galaxies at $<z>$ = 0.34 from the SDSS
%reporting that 20\% of the galaxy light was contained beyond 20 kpc. We agree with
%this result qualitatively. However,
%the authors also stated the central 100 kpc of the galaxy stack could be fit using a single S\'ersic function
%with a r$_{e}$ = 13.1 kpc, and in our case using only one S\'ersic function leaves large residuals 
%(especially in the reddest bands) and none of our galaxies have a effective radius well beyond 5 kpc (for the reddest
%band detections). \citet{Huang13a} fitted 94 of the brightest visually pure ellipticals in the nearby Universe,
%and found that the majority are not well described by a single S\'ersic function. When fitting extra components
%to these galaxies, the authors highlight the low S\'ersic index values for the rest of components. This is also
%found in our sample, modifying the galaxy spheroidal inner parts to a more exponential/less concentrated nature in the outer parts.
our dataset is unique inasmuch as we demonstrate the existence, the relative importance and the spatial distribution 
of this low surface brightness component for each individual galaxy at study.
Of course, longer integration times disclose fainter and fainter features \citep[e.g.][]{Martinez-Delgado10,Duc15,Trujillo16},
which are key to understanding the assembly history of massive galaxies,
although their contribution to the total light and mass decrease in importance. We stress that caution needs to be taken with image data reduction,
as indeed the images must be reduced in such a way to preserve low surface
brightness features. Providing we work in this
direction, the advent of very deep imaging in future years will not only
improve our understanding of high redshift galaxies but will also greatly
enhance our comprehension of the nearby Universe.

We placed constraints on the inside-out growth of massive ETGs
by estimating their observed surface brightness profiles, equivalent Sloan filters restframe profiles and colours, mass profiles and
light cumulative fractions. Both HST bands and the Sloan filters equivalent photometry show a steady decrease in
galaxy flux down to our detection limit without the presence of any truncations. Galaxies displaying signs of merging have
surface brightness bumps in their outer parts (at $>$ 20 kpc; 25-26 mag arcsec$^{-2}$ restframe).
In general, between 20\% and 40\% of the light is located at distances beyond 10 kpc.
Additionally, when comparing
the mean massive ETG mass profiles at different cosmic times, we find that they store a higher fraction
of stellar mass in their outer parts (same galactocentric distances) at decreasing redshift, namely
28.7\% at $< z > =$ 0.1, 15.1\% at $< z > =$ 0.65 and only 3.5\% at $< z > =$ 2.

It is very hard to unambiguously define ETG stellar haloes (especially without kinematic information),
or even comparing with in-situ/accreted material in numerical simulations. However, by integrating both the observational
and simulated mass profiles at distances (10 $<$ R/kpc $<$ 50) where hierarchical accretion is dominant over the in-situ formed stars, we gather evidence for
ETG haloes containing more mass than their late-type counterparts. 
ETG galaxy stellar haloes host 5-20\% of the galaxy mass, in stark
contrast with what has been reported for late-type stellar haloes \citep[see Fig. 12 in][only up to 5\%]{Trujillo16}. 
We must emphasize that the median redshift of the six galaxies at study is $<z> =$0.65, and hence this divergence 
between early- and late-types is larger for local Universe massive ETGs. Extended low
brightness components are present in all massive ETGs in our sample and they seem to be a ubiquitous ingredient of
the $\Lambda$CDM paradigm.

Finally, we developed a toy model in order to attempt to determine the total amount of light and mass
in smooth features linked with ongoing minor merging interactions. Our parametric fits allow us to model the overall spheroid in each galaxy of our ETG sample.
After removing this 2D surface brightness profile, the residual light
gives us insight into the ongoing mass assembly as opposed to more indirect methods such as
satellite counts. The uncertainties are large, due to the necessary assumptions
and the inherent scatter in a galaxy-by-galaxy basis, but the results of this experiment indicate that 
smooth merging features in our imaging contribute at least 1-2\% in galaxy light and mass.
The expectation from close pairs is $\Delta$M/M$\sim$4\%, and our result must be further investigated, 
but it does not contradict the fact major and minor mergers seem to be
the dominant mechanisms driving the evolution of massive ETGs since $z =$ 1.

\section{Acknowledgements}
\label{sec:ack}

We warmly thank the anonymous referee for his thorough reading and questions. 
FB is indebted to James S. Dunlop for his advice, and also for the economic support for MM during her visit to the University of Edinburgh.
The IA Thematic Line \textit{"The assembly history of galaxies resolved in space and time"} is acknowledged for inviting IT for his visit
to the Observatory of Lisbon. We gratefully thank Esther M\'armol-Queralt\'o, Ross McLure, Jes\'us Falc\'on-Barroso, 
Francesco La Barbera, Anton Koekemoer, Hugo Messias and Alexandre Vazdekis and for their help through different stages of this project. Jos\'e
Sabater, Britton Smith and Jovan Veljanoski are very much acknowledged for very valuable computational assistance.
We have extensively used the following software packages: TOPCAT \citep{Taylor05}, ALADIN \citep{Bonnarel2000} and
the IDL routines \texttt{mpfit} and \texttt{mpfitfun} \citep{Markwardt09}.
FB acknowledges the support of the European 
Research Council via the award of an Advanced Grant to James S. Dunlop, the funding from the ASTRODEEP FP7 programme and
the support by FCT via the postdoctoral fellowship SFRH/BPD/103958/2014.
This work is supported by Fundação para a Ciência e a Tecnologia (FCT) through national funds (UID/FIS/04434/2013) and by FEDER through COMPETE2020 (POCI-01-0145-FEDER-007672).
FB and IT also acknowledges support from grant AYA2013-48226-C3-1-P from the Spanish Ministry of Economy and Competitiveness (MINECO).
ECL would like to acknowledge financial support from the ERC via an Advanced Grant under grant agreement no. 321323-NEOGAL.
APC acknowledges a COFUND/Durham Junior Research Fellowship under EU grant [267209].
PGP-G acknowledges support from Spanish Government Grants AYA2012-31277 and AYA2015-70815-ERC.
This work has made use of the Rainbow Cosmological Surveys Database, which is operated by the Universidad Complutense de Madrid (UCM), 
partnered with the University of California Observatories at Santa Cruz (UCO/Lick,UCSC).

\bibliographystyle{mn2e_warrick}
\bibliography{refs}

\appendix
\section{H-band profiles}
\label{appendix:fits}

% irac160002->HUDF-1
% irac160251->HUDF-6
% irac160707->HUDF-5
% irac161317->HUDF-3
% irac159343->HUDF-2
% irac160271->HUDF-4

%\begin{figure*}
%% \begin{center} 
%\vspace{0cm}
%\hspace{2cm}
%\rotatebox{0}{
%\includegraphics[angle=0,width=1.0\linewidth]{szomoru_for_haloes.eps}}
%% \end{center}
%\vspace{-0.5cm}
%
%\caption{Observed (black line), model (convolved and non-convolved with the PSF, coloured solid and dashed lines respectively,
%with colours indicated in the legend) and ``a la Szomoru" (deconvolved adding the residuals of the 4 S\'ersic fit; red line) galaxy surface brightness profiles for 
%our galaxy sample in the H-band. The subplot shows the reduced chi-square ($\chi^{2}_{\nu}$)
%values for the S\'ersic fits we performed. The bottom miniplots display the differences between the observed surface brightness profile
%and the multi-S\'ersic PSF convolved models.}
%\label{fig:fits_szomoru}
%
%\end{figure*}

\begin{figure*}
% \begin{center} 
\centering
\includegraphics[angle=90,trim=-1.cm 0cm 0cm 0cm, clip=false,width=0.49\linewidth]{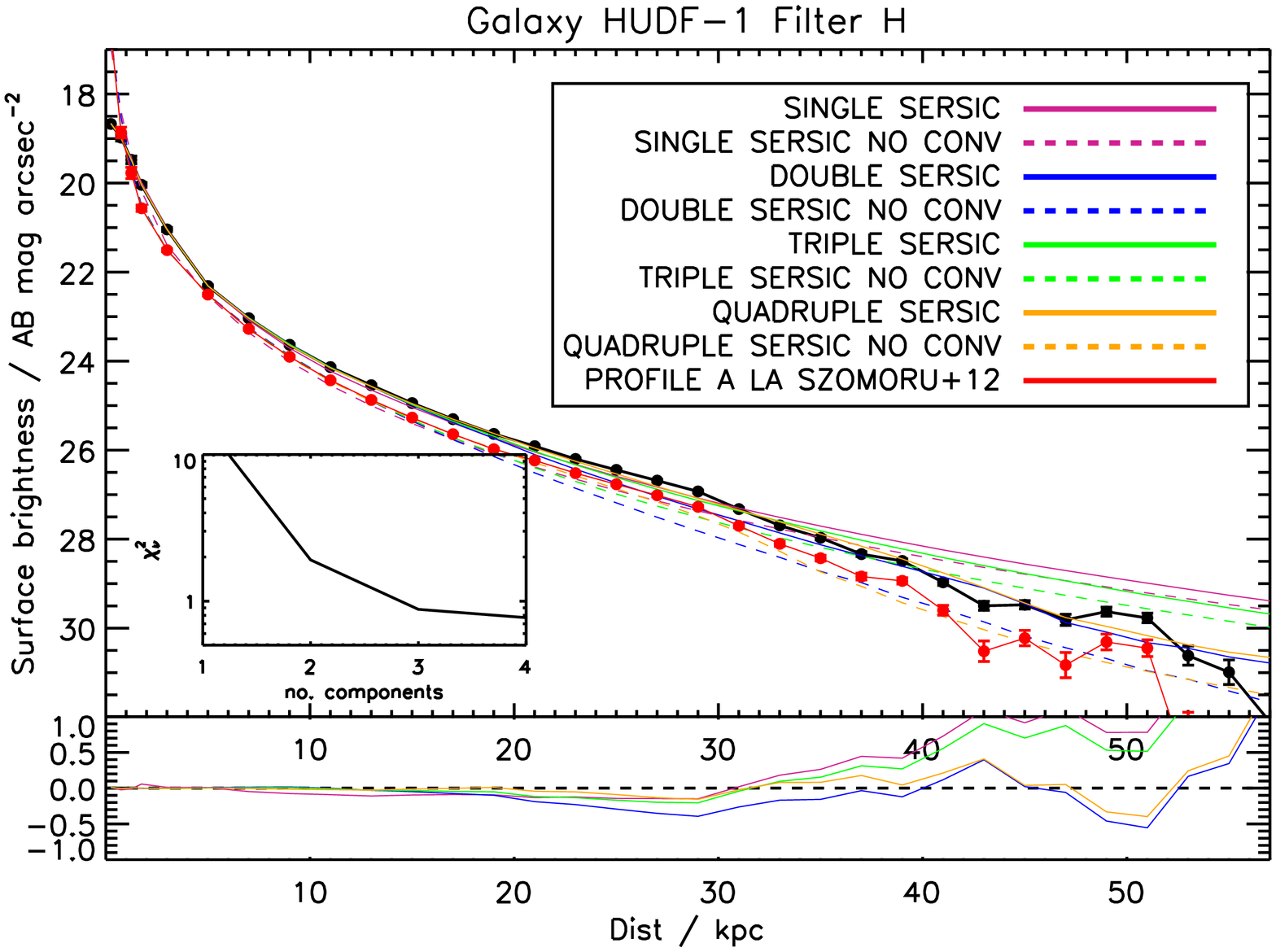} %hudf-1
\includegraphics[angle=90,trim=-1.cm 0cm 0cm 0cm, clip=false,width=0.49\linewidth]{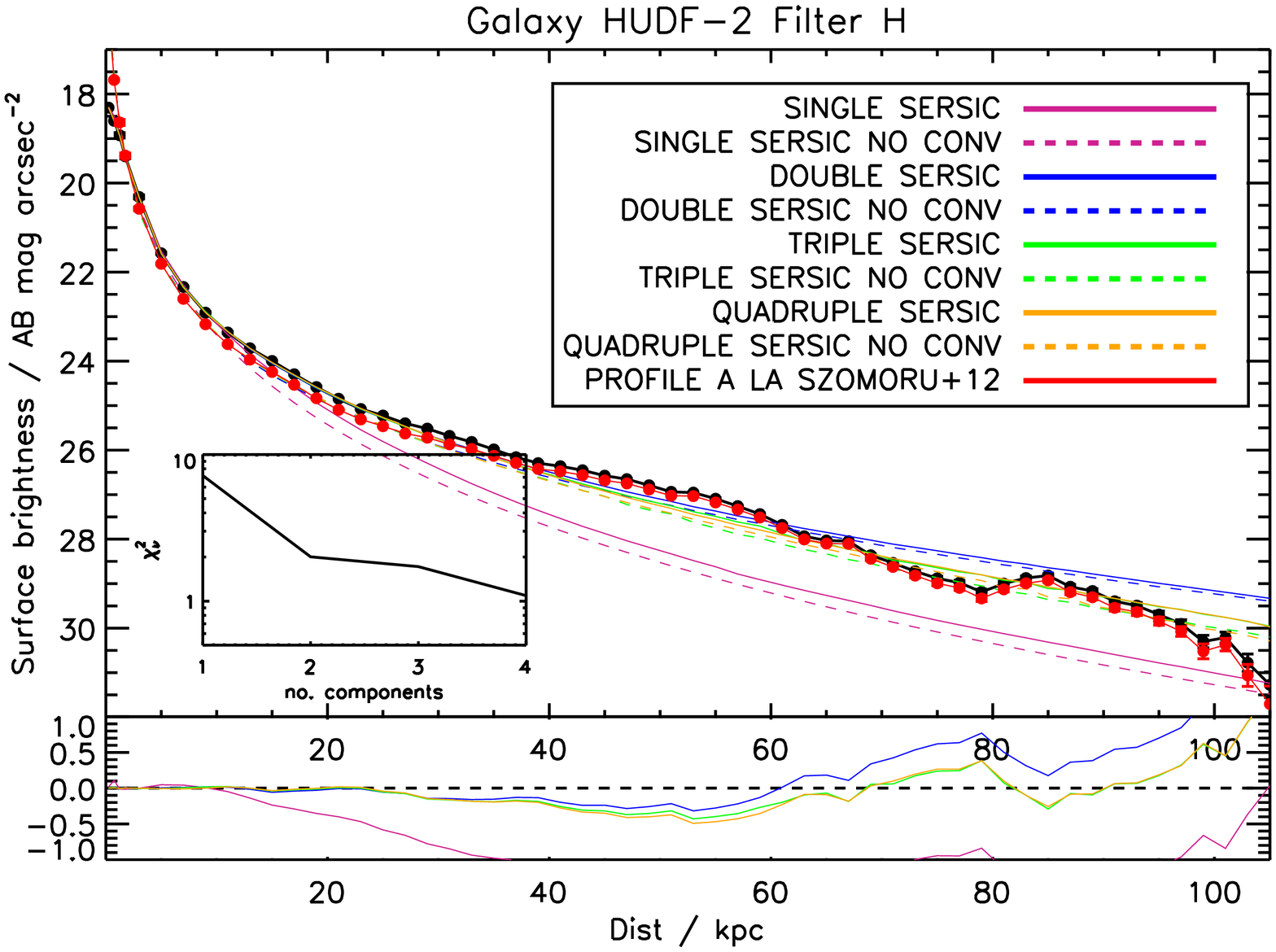} %hudf-2
\includegraphics[angle=90,trim=-1.cm 0cm 0cm 0cm, clip=false,width=0.49\linewidth]{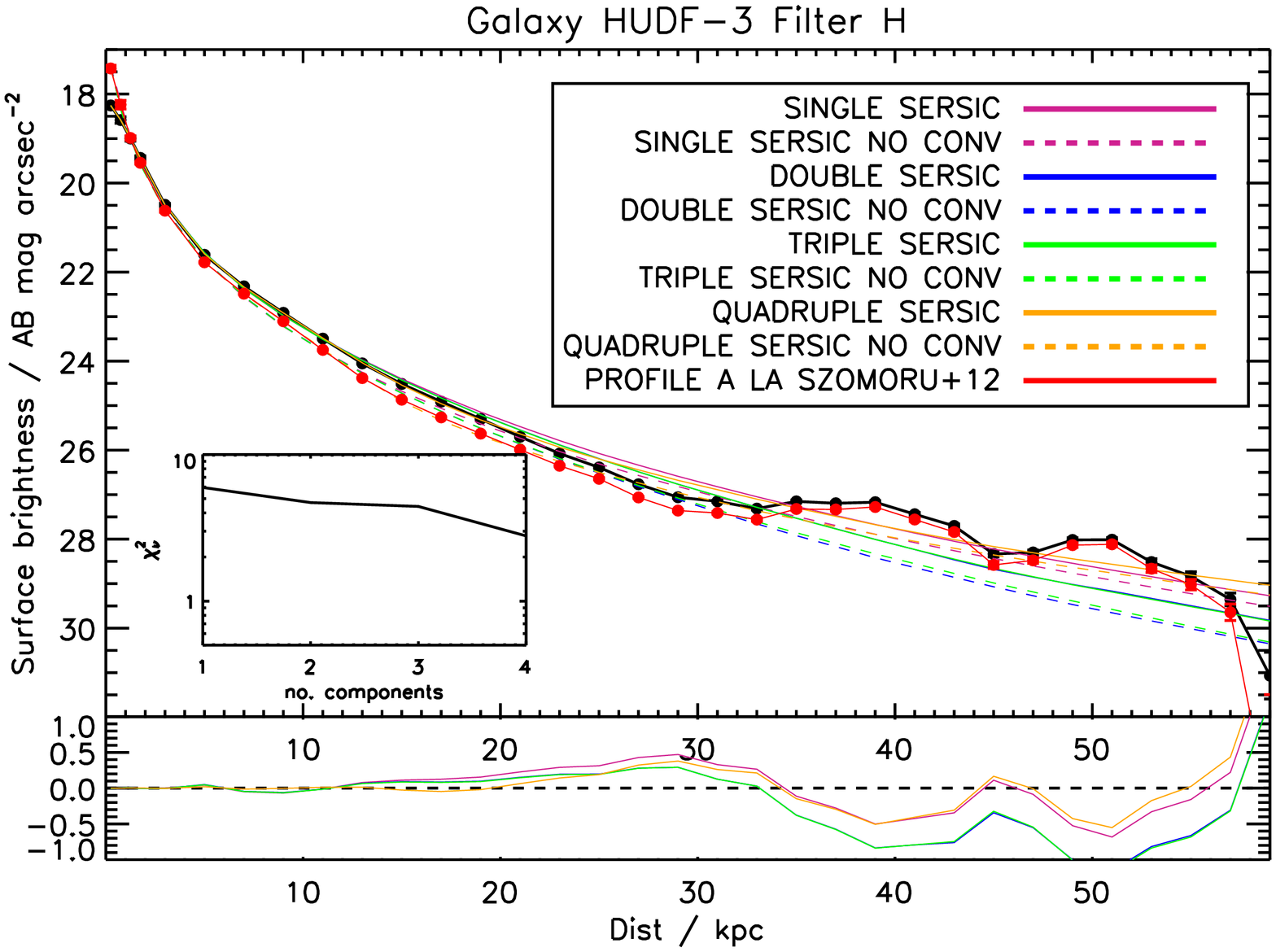} %hudf-3
\includegraphics[angle=90,trim=-1.cm 0cm 0cm 0cm, clip=false,width=0.49\linewidth]{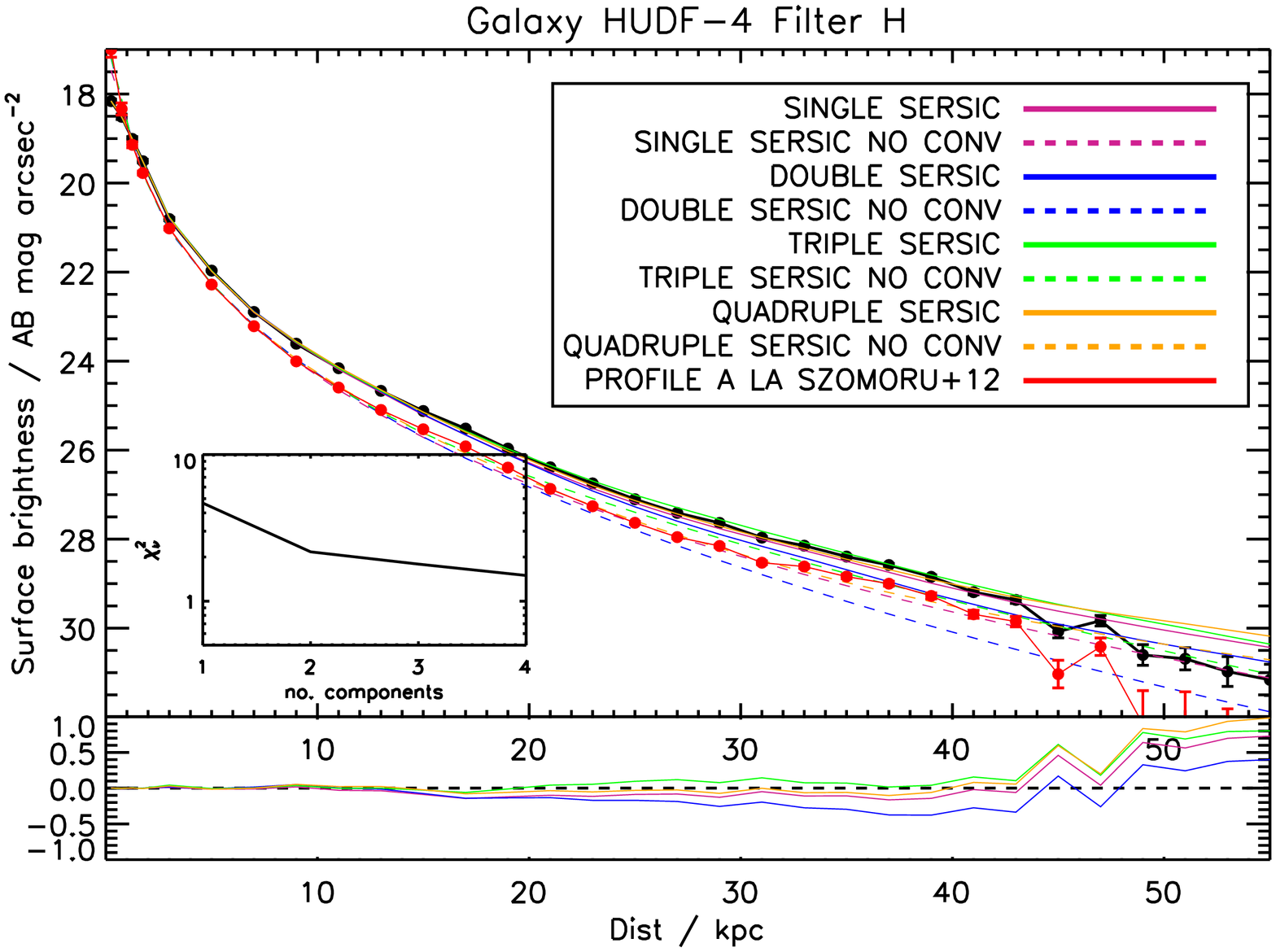} %hudf-4
\includegraphics[angle=90,trim=-2.cm 0cm 0cm 0cm, clip=false,width=0.49\linewidth]{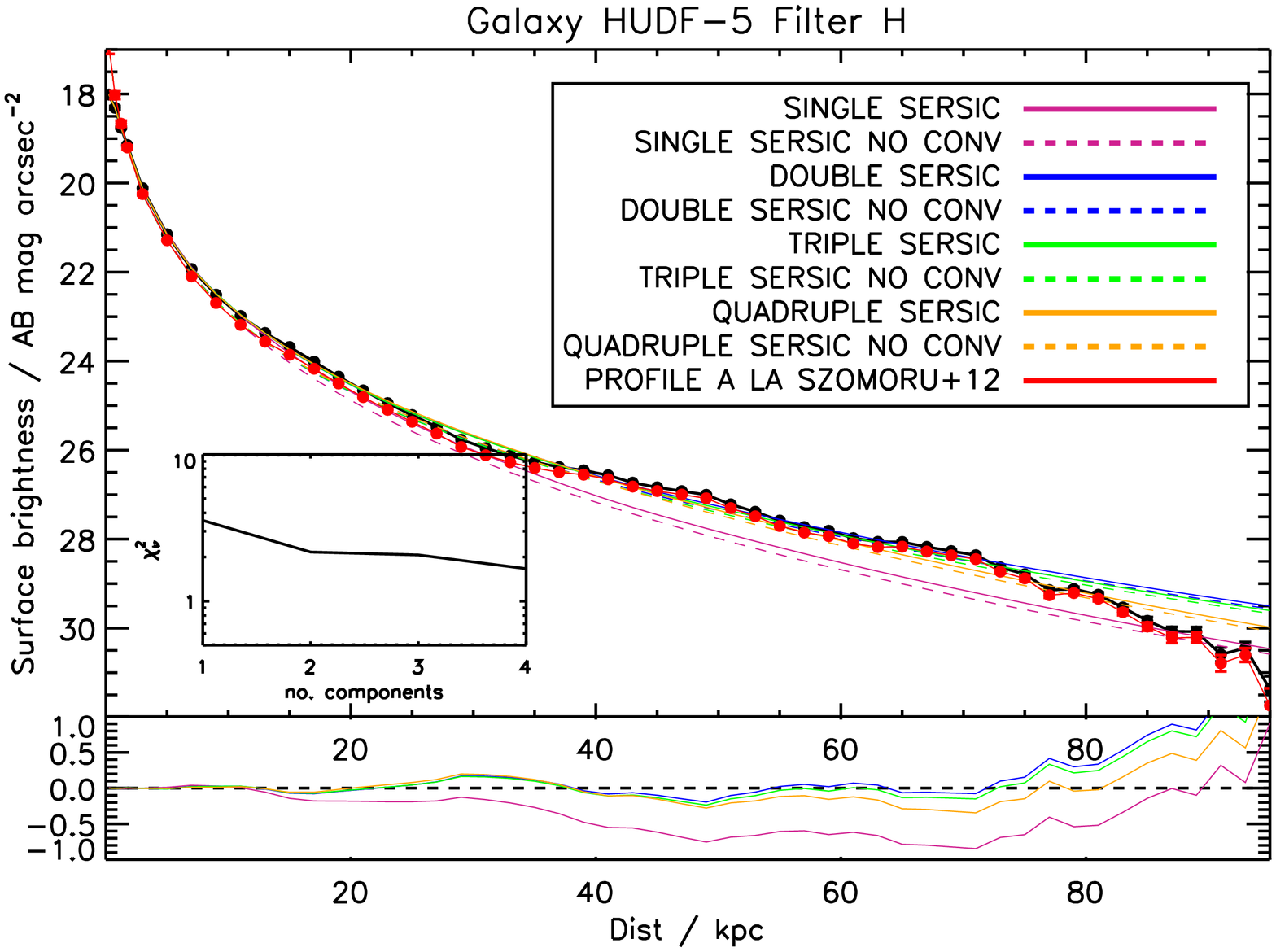} %hudf-5
\includegraphics[angle=90,trim=-2.cm 0cm 0cm 0cm, clip=false,width=0.49\linewidth]{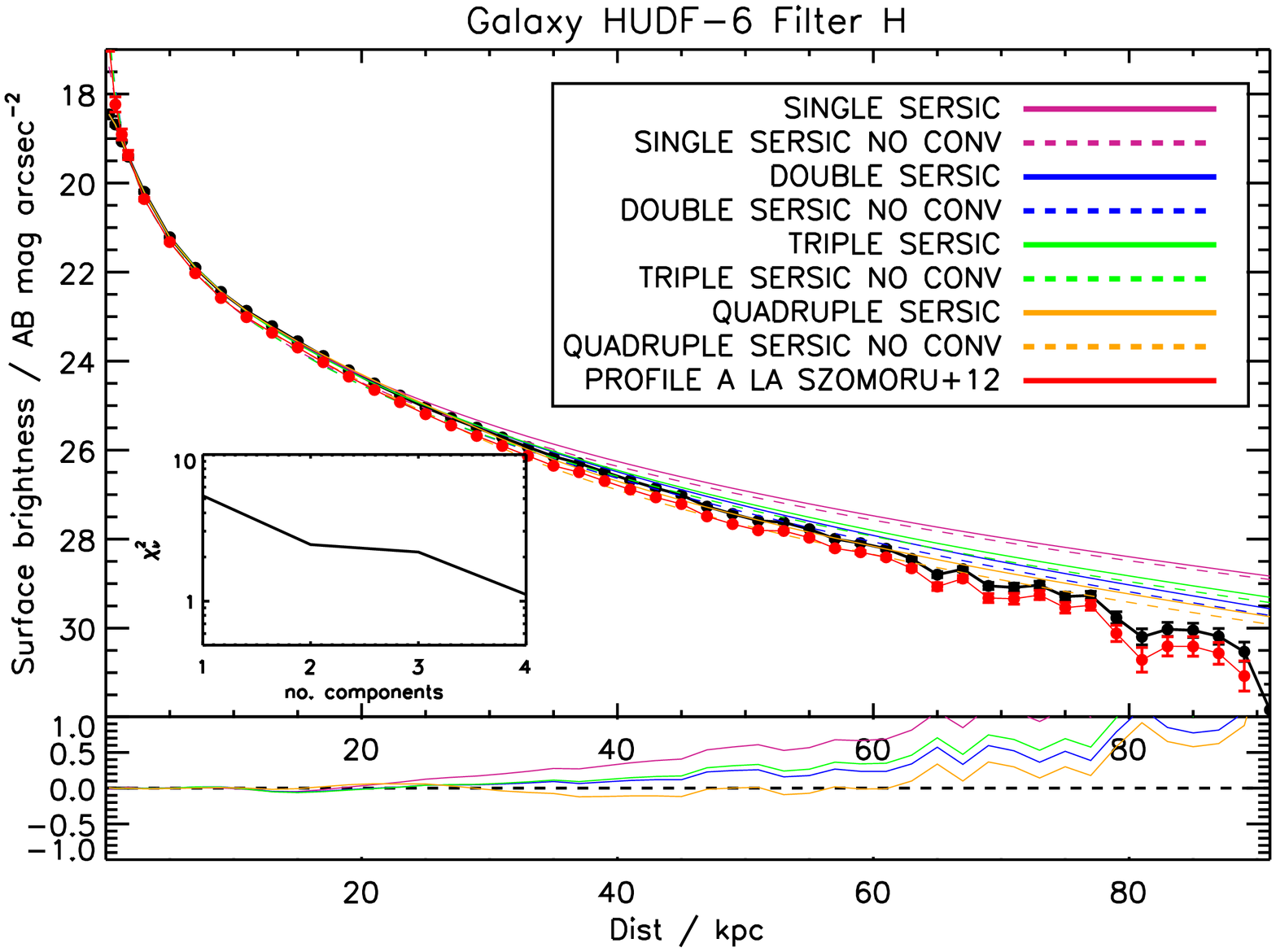} %hudf-6
% \end{center}
\vspace{-0.5cm}

\caption{Observed (black line), model (convolved and non-convolved with the PSF, coloured solid and dashed lines respectively,
with colours indicated in the legend) and ``a la Szomoru" (deconvolved adding the residuals of the 4 S\'ersic fit; red line) galaxy surface brightness profiles for 
our galaxy sample in the H-band. The subplot shows the reduced chi-square ($\chi^{2}_{\nu}$)
values for the S\'ersic fits we performed. The bottom miniplots display the differences between the observed surface brightness profile
and the multi-S\'ersic PSF convolved models.}
\label{fig:fits_szomoru}

\end{figure*}

\end{document}